\documentclass[reprint,prb,showpacs,superscriptaddress, english]{revtex4-1}
\usepackage[utf8]{inputenc}
\usepackage{color}
\usepackage{array}
\usepackage{verbatim}
\usepackage{multirow}
\usepackage{amsmath}
\usepackage{amssymb}
\usepackage{graphicx}
\usepackage{esint}
\usepackage{subscript}
\usepackage[unicode=true,
 bookmarks=false,
 breaklinks=true,pdfborder={0 0 0},backref=false,colorlinks=true]
 {hyperref}

\makeatletter

\providecommand{\tabularnewline}{\\}

\@ifundefined{textcolor}{}
{%
 \definecolor{BLACK}{gray}{0}
 \definecolor{WHITE}{gray}{1}
 \definecolor{RED}{rgb}{1,0,0}
 \definecolor{GREEN}{rgb}{0,1,0}
 \definecolor{BLUE}{rgb}{0,0,1}
 \definecolor{CYAN}{cmyk}{1,0,0,0}
 \definecolor{MAGENTA}{cmyk}{0,1,0,0}
 \definecolor{YELLOW}{cmyk}{0,0,1,0}
}

\usepackage{graphicx}
\usepackage{epstopdf}\usepackage{subfigure}
\usepackage{float}
\usepackage{amsfonts}
\usepackage{bm}% bold math
\usepackage{subfigure}
\usepackage{caption}
\makeatother

\begin{document}

\title{First principle calculation of the effective Zeeman's couplings in topological materials}

\author{Zhida Song}

\affiliation{Beijing National Laboratory for Condensed Matter Physics and Institute
of Physics, Chinese Academy of Sciences, Beijing 100190, China}

\affiliation{University of Chinese Academy of Sciences, Beijing 100049, China}

\author{Song Sun}

\affiliation{Beijing National Laboratory for Condensed Matter Physics and Institute
of Physics, Chinese Academy of Sciences, Beijing 100190, China}

\affiliation{University of Chinese Academy of Sciences, Beijing 100049, China}

\author{Yuanfeng Xu}

\affiliation{Beijing National Laboratory for Condensed Matter Physics and Institute
of Physics, Chinese Academy of Sciences, Beijing 100190, China}

\affiliation{University of Chinese Academy of Sciences, Beijing 100049, China}

\author{Simin Nie}

\affiliation{Department of Materials Science and Engineering, Stanford University,
Stanford, California 94305, USA}

\author{Hongming Weng}

\affiliation{Beijing National Laboratory for Condensed Matter Physics and Institute
of Physics, Chinese Academy of Sciences, Beijing 100190, China}

\affiliation{Collaborative Innovation Center of Quantum Matter, Beijing, 100084,
China}

\author{Zhong Fang}

\affiliation{Beijing National Laboratory for Condensed Matter Physics and Institute
of Physics, Chinese Academy of Sciences, Beijing 100190, China}

\affiliation{Collaborative Innovation Center of Quantum Matter, Beijing, 100084,
China}

\author{Xi Dai}

\email{daix@ust.hk}
\affiliation{Department of Physics, Hong Kong University of Science and Technology, Clear Water Bay, Kowloon 999077, Hong Kong}

\affiliation{Beijing National Laboratory for Condensed Matter Physics and Institute
of Physics, Chinese Academy of Sciences, Beijing 100190, China}

\begin{abstract}
In this paper, we propose a first principle calculation method for the effective Zeeman's coupling
 based on the second perturbation theory and apply it to a few topological materials.
For Bi and Bi$_2$Se$_3$, our numerical results are in good accord with the experimental data;
 for Na$_3$Bi, TaN, and ZrTe$_5$, the structure of the multi-bands Zeeman’s couplings are discussed.
Especially, we discuss the impact of Zeeman's coupling on the Fermi surface's topology in Na$_3$Bi in detail.
\end{abstract}
\maketitle

\section{Introduction}

The Landé g-factor, or the form of effective Zeeman effect in particular materials, has been an old topic in the condensed matter physics, which determines how a pair of otherwise degenerate Kramas states split under the external magnetic field and can lead to a series of consequences in magnetic responses such as the Van Vleck paramagnetism in insulators, the Pauli paramagnetism in metals \cite{ashcroft_solid_book}, as well as the frequency splitting of quantum oscillations in metals or doped semiconductors \cite{oscillation_book}.
The widely used theory about the origin of the effective Landé g-factor in materials was developed in 1950s by Luttinger \cite{luttinger_quantum_1956} and Cohen \cite{cohen_g-factor_1960}, according to which, the Zeeman's coupling is not only contributed by the spin momentums but also the orbital momentums of the Bloch states. As clarified by Chang and Niu in a semiclassical picture, such an orbital contribution can be interpreted as the self-rotating effect of the quasi-particle wave packet under an external magnetic field \cite{chang_berry_1996,chang_berry_2008}.
However, although the theory and physical picture of the effective Zeeman's coupling have been estabilished for a long time, there are still no reported first principle calculations based on the density functional theory (DFT), which motivates us to make a first try. In this work, we present a successful first principle realization of the effective Zeeman's coupling parameters within the projector augmented-wave (PAW) formulation of the DFT.

In the traditional semiconductors described by the Fermi liquid theory, the responses of the system to the external magnetic field can be well described by two sets of coefficients, the effective mass and Landé g-factor tensors. 
In traditional metals and semi-conductors, the Zeeman effect only splits the two degenerate states and modifies the Landau level behavior of the free electrons under an external magnetic field. While, for topological materials with strong SOC the Zeeman effect will also generate otherwise fully compensated Berry's curvature near the Fermi energy, which will affect various of transport properties of the materials.
\iffalse In contrast, in the emerging novel topological materials, where new types of quasi-particle excitations such as gapless Dirac fermions \cite{BiO2, wang_Na3Bi, nagaosa_dirac_nc, Na3Bi_exp}, Weyl fermions \cite{burkov_weyl, wan_weyl, weng_TaAs,huang_TaAs, TaAs_exp_iop, TaAs_exp_Hasan}, nodal line fermions \cite{burkov_nodal,weng_nodal,kim_nodal,fang_nodal,nodal_Cu3PdN, nodal_PbTaSe2, nodal_chain} and the new fermions \cite{bradlyn_new_fermion, weng_TaN, MoP_exp}, etc. dominate the low energy physics, new ingredients including the Berry phase \cite{niu_RMP, son_berry_prl, chiral_kinetic_thoery} and the multi-bands Zeeman's coupling, as discussed in this paper, may play important roles in magnetic responses. \fi 
The consequences of the Berry curvature have been studied thoroughly in various theoretical and experimental works, including the weak localization \cite{lu_weakloc}, the negative magnetoresistance \cite{abj_1983,burkov_CA,burkov_NMR,son_NMR_2013,Na3Bi_NMR}, the nonlocal transport \cite{CA_nonlocal}, the anomalous quantum oscillation \cite{luhz_SdH_2016}, and the anomalous coupling with pseudo scalar phonon \cite{song_detecting_2016}, etc.
However, the role of the Zeeman's coupling in these transport phenomenons has not attracted much attentions until recently \cite{luhz_MR_TI} , which makes it timely and important to develop the first principle method to calculate the g-factor tensor for realistic material systems.

The paper is organized as the following. In section \ref{sec:Theory}, the theory of effective Zeeman's coupling will be introduced in the framework of the second order quasi-degenerate perturbation theory. In section \ref{sec:results}, the first principle calculation method is introduced and applied to a few topological materials.
In section \ref{sec:discussion}, a most direct application of our data, i.e. the Fermi surface topology of Dirac semimetal under Zeeman's coupling, is discussed.
In the end, we give a brief summary in section \ref{sec:summary}.

\section{Theory\label{sec:Theory}}

In this section, we will give a short review about the quasi-degenerate perturbation theroy \cite{lowdin1951note,luttinger_perturbation,winkler_spin-orbit_2003} and show how the effective Zeeman's coupling emerges as a gauge invariant second order perturbation.

In semiconductors and semimetals the low energy physics usually involves only a small part of the Brillouin zone, i.e. the neighbourhoods around a few special wave vectors, and so can be well described by a few k$\cdot$p Hamiltonians around these points. 
For each of the wave vectors $\mathbf{K}$, we can write the k$\cdot$p Hamiltonian $e^{-i\left(\mathbf{K}+\mathbf{k}\right)\cdot\mathbf{r}}\hat{H}e^{i\left(\mathbf{K}+\mathbf{k}\right)\cdot\mathbf{r}}$
on the periodic parts of Bloch wave functions as \cite{winkler_spin-orbit_2003}
\begin{align*}
H_{nn^{\prime}}\left(\mathbf{k}\right) & =\delta_{nn^{\prime}}\left[\epsilon_{n}+\frac{\hbar^{2}\mathbf{k}^{2}}{2m_{e}}\right]+\frac{\hbar}{m_{e}}\boldsymbol{\pi}_{nn^{\prime}}\cdot\mathbf{k}
\end{align*}
Here $\epsilon_{n}$ is the $n$-th band energy at $\mathbf{K}$,
and $\boldsymbol{\pi}_{nn^{\prime}}=\left\langle \psi_{n\mathbf{K}}\right|\hat{\mathbf{p}}+\frac{1}{2m_{e}c^{2}}\left(\hat{\mathbf{s}}\times\nabla V\right)\left|\psi_{n^{\prime}\mathbf{K}}\right\rangle $
is the momentum element with spin-orbital coupling (SOC) correction, with $\hat{\mathbf{p}}$ being the canonical momentum operator, $\hat{\mathbf{s}}$ being the spin operator, $V$ being the scalar potential in crystal, $m_{e}$ being the electron mass, $c$ being the light speed, and $\left|\psi_{n^{\prime}\mathbf{K}}\right\rangle $ being the $n^{\prime}$-th bloch wave function. 
Since the excitations near the Fermi level dominate the low energy physics, we only include a small number of bands near the Fermi level. However, a direct cutoff in the Hilbert space is a too rough approximation because the high energy subspace couples to the low energy subspace through the off-diagonal Hamiltonian. To solve this problem, in the quasi-degenerate perturbation theory the coupling Hamiltonian is treated as a small quantity and an unitary transformation is constructed in series of it to decouple the two subspaces. With the second order approximation, the transformed Hamiltonian in the low energy subspace can be derived as
\begin{align}
 & H_{mm^{\prime}} =\delta_{mm^{\prime}}\left[\epsilon_{m}+\frac{\hbar^{2}\mathbf{k}^{2}}{2m_{\mathrm{e}}}\right]+\frac{\hbar}{m_{\mathrm{e}}}\boldsymbol{\pi}_{mm^{\prime}}\cdot\mathbf{k}\nonumber \\
& + \frac{\hbar^{2}}{2m_{\mathrm{e}}^{2}}\sum_{l}^{\prime}\sum_{ij}\left[\frac{1}{\epsilon_{m}-\epsilon_{l}}+\frac{1}{\epsilon_{m^{\prime}}-\epsilon_{l}}\right]\pi_{ml}^{i}\pi_{lm^{\prime}}^{j}k^{i}k^{j}
\end{align}
, where $m,m^{\prime}$ are the band indexes in the low energy subspace, and the summation over $l$ is limited within the high energy subspace.
In the presence of magnetic field, according to the Peierls substitution the momentums $\hbar k^{i}$ should be replaced by the kinetic momentum operators $-i\hbar\partial^{i}-eA^{i}$, which are not commutative with each other. Thus the product $\left(-i\hbar\partial^{i}-eA^{i}\right)\left(-i\hbar\partial^{j}-eA^{j}\right)$ substituting $\hbar^{2}k^{i}k^{j}$ can be decomposed into a gauge dependent symmetric component $\frac{1}{2}\left\{ -i\hbar\partial^{i}-eA^{i},-i\hbar\partial^{j}-eA^{j}\right\} $ and a gauge invariant anti-symmetric component $\frac{1}{2}\left[-i\hbar\partial^{i}-eA^{i},-i\hbar\partial^{j}-eA^{j}\right]$ which equals to $-\frac{i\hbar e}{2}\sum_{k}\epsilon^{ijk}B^{k}$ \cite{luttinger_quantum_1956,cohen_g-factor_1960,Yuki_gfactor}. 
Here $A^{i}$ is the vector potential, $\epsilon^{ijk}$ is the Levi-Civita tensor, and, the curly and square brakets represent the anti-commutator and the commutator, respectively. Finally, the total Hamiltonian in magnetic field can be summarized as a gauge dependent part
\begin{align}
  \hat{H}_{mm^{\prime}}^{kp} &=\delta_{mm^{\prime}}\epsilon_{m}+\frac{\hbar}{m_{\mathrm{e}}}\boldsymbol{\pi}_{mm^{\prime}}\cdot\left(-i\nabla+\frac{e}{\hbar}\mathbf{A}\right)\nonumber \\
 & +\sum_{ij}M_{mm^{\prime}}^{ij}\left(-i\partial^{i}+\frac{e}{\hbar}A^{i}\right)\left(-i\partial^{j}+\frac{e}{\hbar}A^{j}\right)\label{eq:Hkp}
\end{align}
, and a gauge invariant part, i.e. the effective Zeeman's coupling
\begin{equation}
\hat{H}_{mm^{\prime}}^{Z}=\mu_{\mathrm{B}}\frac{1}{\hbar}\left(\mathbf{L}_{mm^{\prime}}+2\mathbf{s}_{mm^{\prime}}\right)\cdot\mathbf{B}\label{eq:HZ}
\end{equation}
Here $M_{mm^{\prime}}^{ij}$ is the symmetrized rank-2 tensor describing the inverse effective mass
\begin{align}
M_{mm^{\prime}}^{ij} & =\delta_{mm^{\prime}}\delta_{ij}\frac{\hbar^{2}}{2m_{\mathrm{e}}}+\frac{\hbar^{2}}{2m_{\mathrm{e}}^{2}}\sum_{l}^{\prime}\left[\frac{1}{\epsilon_{m}-\epsilon_{l}}+\frac{1}{\epsilon_{m^{\prime}}-\epsilon_{l}}\right]\nonumber \\
 & \times\frac{\pi_{ml}^{i}\pi_{lm^{\prime}}^{j}+\pi_{ml}^{j}\pi_{lm^{\prime}}^{i}}{2}\label{eq:M}
\end{align}
, $\mathbf{L}_{mm^{\prime}}$ is the effective orbital momentum contributed by the anti-symmetrized rank-2 tensor as
\begin{equation}
L_{mm^{\prime}}^{k}=\frac{-i\hbar}{2m_{\mathrm{e}}}\sum_{l}^{\prime}\sum_{ij}\left[\frac{1}{\epsilon_{m}-\epsilon_{l}}+\frac{1}{\epsilon_{m^{\prime}}-\epsilon_{l}}\right]\epsilon_{ijk}\pi_{ml}^{i}\pi_{lm^{\prime}}^{j}\label{eq:L}
\end{equation}
, $\mu_{\mathrm{B}}=\frac{e\hbar}{2m_{\mathrm{e}}}$ is the Bohr magneton, and $\frac{2}{\hbar}\mu_{B}\mathbf{s}\cdot\mathbf{B}$ is the bare Zeeman's coupling from Schr{\"o}dinger's equation.

In the above derivation, we would like to emphasize two points. The first point is that both the form and value of the g factor tensor themselves are not the observable quantities because it depends on the low energy sub-space chosen by us. Therefore, in principle only the resulting final Landau level spectrum or other transport properties are measurable quantities. The second point is that such a quasi-degenerate perturbation theory is well defined only if there is a relatively large gap, compared to other energy scales such as Fermi energy, between the low and high energies energy sub-spaces.

\section{First principle calculations \label{sec:results}}

In this section, we calculate the second order k$\cdot$p models and the effective Zeeman's couplings for a few typical topological materials by Eq. (\ref{eq:M}) and (\ref{eq:L}), in which the momentum and spin matrix elements are computed with the Vienna ab-initio simulation package(VASP). Technical details about the computation method within the projector augmented-wave (PAW) formulation are given in appendix \ref{sec:PAW}. The exchange and correlation potential is treated within some most widely used types, i.e. the local density approximation (LDA) \cite{Kohn_Sham,LDA_CA}, the generalized gradient approximation (GGA) \cite{perdew_GGA}, and the LDA correlation plus modified Becke and Johnson (mBJ) exchange potential\cite{mBJ_2009}, and a bands cutoff of 300 is choosen for the summation
over $l$ in Eq. (\ref{eq:M}) and (\ref{eq:L}).

\subsection{Two bands models\label{sub:2bands}}

The g factor of Schr{\"o}dinger electron is a dimensionless scalar that characterizes the ratio between electron's magnetic momentom and spin momentum, which equals to 2 in vaccum. 
In semiconductors where both inversion and time reversal (TR) symmetry are present, the conduction or valance bands are doublely dengenerate and are well described by a quasi-Schr{\"o}dinger's equation with renormalized mass and g factor. Generally speaking, according to the anistropy that may occur in real materials, both the effective mass and effective g factor should be 3 by 3 tensors now. 
Here we define the effective g factor tensor as the expansion coefficients of the Zeeman's coupling Hamiltonian on Pauli's matrices
\begin{equation}
H_{mm^{\prime}}^{\mathrm{Z}}=\mu_{\mathrm{B}}\frac{1}{2}\sum_{i}g_{ij}\sigma_{mm^{\prime}}^{j}B^{i}
\end{equation}
, which is reduced to $g_{ij}=2\delta_{ij}$ in vaccum.

In the following, we will present two examples, i.e. bismuth and Bi$_2$Se$_3$ whose g factors have been measured in experiments \cite{verdun_far-infrared_1976,gfactor_Bi2Se3} or studied in previous theoretical work \cite{Yuki_gfactor}, to verify the validity of our method.

\subsubsection{Hole pocket in Bi}

Bismuth has played an important role in the topological material family because of its large SOC, which, in Ref. [\onlinecite{alicea_bismuth_2009}], is also believed to be responsible for the large ansitropic Zeeman's coupling of the valence bands in its elemental crystal. The elemental bismuth has the space group $R\bar{3}m$ and is a typical semimetal with a hole pocket at T and three equivalent electron pockets at L\cite{liu_electronic_1995}. Here we only focus on the hole pocket bands, which form a two dimensional IR $E_{\frac{3}{2}u}$ of the little group $D_{3d}$ at T due to the TR symmetry. Choosing the two bases as $|P\frac{3}{2}\rangle$, $|P\bar{\frac{3}{2}}\rangle$, we find that all the mass and g factor elements, except $m_{xx}=m_{yy}=m_{\perp}$, $m_{zz}=m_{\parallel}$, $g_{zz}=g_{\parallel}$, are zero. In fact, the absence of $g_{xx}$ and $g_{yy}$, or, the absence of off-diagonal elements in $H^{Z}$, is guaranteed by the angular momentum conservation condition $\langle P\frac{3}{2}|L_{x}\pm iL_{y}|P\bar{\frac{3}{2}}\rangle=0$, which indicates that the Zeeman effect for the in-plane field is nonlinear.
Numerical results are summarized in table (\ref{tab:2b}). It shows that the experimental g factors and masses lie between the LDA/GGA and mBJ values, which is very reasonable because LDA and GGA usually underestimate the band gap while mBJ may overestimate the band gap.

\subsubsection{Bi\protect\textsubscript{2}Se\protect\textsubscript{3}}

The second system we study is $\mathrm{Bi_{2}Se_{3}}$ with the space group $R\bar{3}m$, which has been a famous material in the past few years as the first large gap three dimensional $\mathbb{Z}_{2}$ topological insulator \cite{zhang_Bi2Se3,xueqk_Bi2Se3}. Besides its non-trivial topology, $\mathrm{Bi_{2}Se_{3}}$ also has non-trivial large g factors in its conduction bands, which are doublely degenerate and form the irreducible representation (IR) $E_{\frac{1}{2}u}$ of the little group $D_{3d}$ at $\Gamma$ \cite{gfactor_Bi2Se3}. Taking the gauge in which the bases transform as $|P\frac{1}{2}\rangle$ and $|P\bar{\frac{1}{2}}\rangle$ (see Ref. [\onlinecite{altmann_point-group}]), we find that due to the symmetry constraints the effective mass and g factor tensor have very concise forms: $m_{xx}=m_{yy}=m_{\perp}$, $m_{zz}=m_{\parallel}$, $g_{xx}=g_{yy}=g_{\perp}$, $g_{zz}=g_{\parallel}$, and all the other components are zero.
Results comparable with experimental data are obtained, as shown in table (\ref{tab:2b}), which again shows that the experimental values lie between the LDA/GGA and mBJ values.
An interesting observation is that the mBJ gap here is smaller than the LDA/GGA gap, which is different with the tendency in bismuth, because of the band inversion nature in its band structure.
%An interesting observation is that the tendency of the band gap
% with respect to the exchange and correlation potential here
% is opposite with the tendency in bismuth, i.e. the mBJ gap is smaller
% than the LDA/GGA gap, because of the band inversion.

\begin{table}
\begin{centering}
\begin{tabular}{|c|c|cccc|}
\hline
\multicolumn{2}{|c|}{} & \multicolumn{1}{c|}{LDA} & \multicolumn{1}{c|}{GGA} & \multicolumn{1}{c|}{mBJ} & Exp\tabularnewline
\hline
\multirow{4}{*}{$\mathrm{Bi}$ hole} & Gap(eV) & 0.272 & 0.298 & 0.453 & 0.18$\sim$0.41 \cite{liu_electronic_1995}\tabularnewline
\cline{2-6}
 & $g_{\parallel}$ & 79.40 & 73.83 & 55.31 & 63.2 \cite{verdun_far-infrared_1976}\tabularnewline
\cline{2-6}
 & $m_{\parallel}^{*}$ & -0.592 & -0.604 & -0.626 & -0.69$\sim$-0.702 \cite{liu_electronic_1995}\tabularnewline
\cline{2-6}
 & $m_{\perp}^{*}$ & -0.0389 & -0.0447 & -0.0895 & -0.064 \cite{liu_electronic_1995}\tabularnewline
\hline
\multirow{5}{*}{$\mathrm{Bi_{2}Se_{3}}$} & Gap(eV) & 0.521 & 0.435 & 0.236 & \tabularnewline
\cline{2-6}
 & $g_{\parallel}$ & 18.4 & 21.76 & 41.80 & 32 \cite{gfactor_Bi2Se3}\tabularnewline
\cline{2-6}
 & $g_{\perp}$ & 16.37 & 17.86 & 26.18 & 23 \cite{gfactor_Bi2Se3}\tabularnewline
\cline{2-6}
 & $m_{\parallel}^{*}$ & 0.851 & 0.488 & 0.238 & \tabularnewline
\cline{2-6}
 & $m_{\perp}^{*}$ & 2.61 & 0.420 & 0.077 & 0.124 \cite{gfactor_Bi2Se3}\tabularnewline
\hline
\end{tabular}
\par\end{centering}

\protect\caption{\label{tab:2b}Effective g factor and mass tensors for the two bands models of conduction bands at $\Gamma$ of $\mathrm{Bi_{2}Se_{3}}$ and valance bands at T of $\mathrm{Bi}$. Both $g$ and $m^{*}$ are dimensionless numbers. The energy gaps shown here are the direct gaps at these high symmetry points.}
\end{table}

\subsection{Four bands models}

In topological semimetals with gapless nodes or topological insulators near critical point, due to the zero or exremely small band gap, there is no longer a well defined second order quasi-degenerate perturbation theory for a two bands subspace. Therefore the effective model should include both the conduction and valley bands, and consequently the concept of g factor tensor is no longer valid and a multi-bands Zeeman's coupling has to be considered. We will show that, such a multi-bands Zeeman's coupling has much more fruitful physical consequences.

\subsubsection{Na\protect\textsubscript{3}Bi}

A typical instance of Dirac semimetal is $\mathrm{Na_{3}Bi}$ with space group $P6_{3}/mmc$, where the two Dirac nodes are generated by the crossings of two doubly degenerate bands, i.e. the $\pm\frac{3}{2}$ and $\pm\frac{1}{2}$ states forming the $E_{\frac{3}{2}}$ and $E_{\frac{1}{2}}$ IRs of the little group $C_{6v}$ along $z$ axis \cite{wang_Na3Bi,Na3Bi_exp}.
Here the little group $C_{6v}$ is crucial for the existance of Dirac nodes, as the crossing between different IRs is protected by the rotational symmetry and the double degeneracy of each IR is gauranteed by the vertical mirrors. Expand the k$\cdot$p Hamiltonian around one of the Dirac node, say, $\left(00k_{c}\right)$ for example, we get the effective k$\cdot$p model as
\begin{align}
 & H^{kp}\left(\mathbf{k}\right)=C\left(\mathbf{k}\right)+\nonumber \\
 & \begin{bmatrix}-M\left(\mathbf{k}\right) & -v_{\perp}k_{-}-\gamma_{1}k_{z}k_{-} & \gamma_{2}k_{-}^{2} & 0\\
* & M\left(\mathbf{k}\right) & 0 & \gamma_{2}k_{-}^{2}\\
* & * & M\left(\mathbf{k}\right) & v_{\perp}k_{-}+\gamma_{1}k_{z}k_{-}\\
* & * & * & -M\left(\mathbf{k}\right)
\end{bmatrix}\label{eq:Hk-Na3Bi}
\end{align}
and the effective Zeeman's coupling as
\begin{equation}
H^{Z}=\mu_{B}\begin{bmatrix}g_{\parallel}^{\frac{3}{2}}B_{z} & g_{\perp}^{\prime}B_{-} & 0 & 0\\
* & g_{\parallel}^{\frac{1}{2}}B_{z} & g_{\perp}^{\frac{1}{2}}B_{-} & 0\\
* & * & -g_{\parallel}^{\frac{1}{2}}B_{z} & g_{\perp}^{\prime}B_{-}\\
* & * & * & -g_{\parallel}^{\frac{3}{2}}B_{z}
\end{bmatrix}\label{eq:HZ-Na3Bi}
\end{equation}
Here the basis set is choosen as $|\frac{3}{2}\rangle$, $|\frac{1}{2}\rangle$,
$|\bar{\frac{1}{2}}\rangle$, $|\bar{\frac{3}{2}}\rangle$, and the quantities in above equations are defined as $k_{\pm}=k_{x}\pm ik_{y}$, $B_{\pm}=B_{x}\pm iB_{y}$, $M\left(\mathbf{k}\right)=v_{z}k_{z}+M_{\parallel}k_{z}^{2}+M_{\perp}\left(k_{x}^{2}+k_{y}^{2}\right)$, and $C\left(\mathbf{k}\right)=v_{0}k_{z}+C_{\parallel}k_{z}^{2}+C_{\perp}\left(k_{x}^{2}+k_{y}^{2}\right)$, respectively. 
As shown in Eq. (\ref{eq:HZ-Na3Bi}), such a matrix form Zeeman's coupling not only splits each Weyl subblock but also couples the two Weyl subblocks together, leading to some exotic Fermi surface structure, as will be discussed in the next section. The calculated model parameters are summarized in table (\ref{tab:4b}).

\begin{table*}
\begin{centering}
\begin{tabular}{|cccc|cccc|cccc|cccc|c|}
\hline
\multicolumn{4}{|c|}{$\mathrm{Na_{3}Bi}$} & \multicolumn{4}{c|}{$\mathrm{TaN}$} & \multicolumn{4}{c|}{$\mathrm{ZrTe}_{5}$} & \multicolumn{4}{c|}{$\mathrm{Bi}$ electron} & \multirow{2}{*}{Unit}\tabularnewline
\cline{1-16}
\multicolumn{1}{|c|}{} & \multicolumn{1}{c|}{LDA} & \multicolumn{1}{c|}{GGA} & mBJ & \multicolumn{1}{c|}{} & \multicolumn{1}{c|}{LDA} & \multicolumn{1}{c|}{GGA} & mBJ & \multicolumn{1}{c|}{} & \multicolumn{1}{c|}{LDA} & \multicolumn{1}{c|}{GGA} & mBJ & \multicolumn{1}{c|}{} & \multicolumn{1}{c|}{LDA} & \multicolumn{1}{c|}{GGA} & mBJ & \tabularnewline
\hline
$g_{\parallel}^{\frac{3}{2}}$ & 5.78 & 5.84 & 6.36 & $g_{\parallel}^{\frac{3}{2}}$ & -0.25 & -0.28 & -0.36 & $g_{x}^{p}$ & -0.12 & -0.04 & 0.08 & $g_{x}^{p}$ & -2.73 & -2.70 & -2.56 & \multirow{6}{*}{1}\tabularnewline
$g_{\parallel}^{\frac{1}{2}}$ & 2.90 & 2.96 & 3.41 & $g_{\parallel}^{\frac{1}{2}}$ & 1.84 & 1.96 & 2.45 & $g_{y}^{p}$ & 12.24 & 11.63 & 9.66 & $g_{y}^{p}$ & 5.90 & 5.94 & 5.11 & \tabularnewline
$g_{\perp}^{\frac{1}{2}}$ & 4.40 & 4.45 & 4.50 & $g_{\perp}^{\frac{1}{2}}$ & 0.54 & 0.57 & 0.66 & $g_{z}^{p}$ & -5.19 & -4.56 & -2.22 & $g_{z}^{p}$ & 1.08+0.29i & 1.05+0.25i & 0.55+0.19i & \tabularnewline
$g_{\perp}^{\prime}$ & 3.10 & 3.12 & 3.06 & $g_{\perp}^{\prime}$ & 0.25 & 0.23 & 0.16 & $g_{x}^{s}$ & 0.64 & 0.67 & 0.77 & $g_{x}^{s}$ & 0.025 & 0.054 & 0.033 & \tabularnewline
 &  &  &  &  &  &  &  & $g_{y}^{s}$ & -2.42 & -2.89 & -6.45 & $g_{y}^{s}$ & 5.98 & 5.99 & 5.98 & \tabularnewline
 &  &  &  &  &  &  &  & $g_{z}^{s}$ & -0.55 & -0.61 & -0.93 & $g_{z}^{s}$ & 2.64-1.98i & 2.60-1.98i & 1.83-2.47i & \tabularnewline
\hline
 &  &  &  & $\Delta$  & 2.74 & 2.68 & 2.48 & $\Delta$  & 0.0113 & 0.0118 & 0.0323 & $\Delta$ & 0.113 & 0.102 & 0.013 & $\mathrm{eV}$\tabularnewline
\hline
$v_{0}$  & 1.41 & 1.47 & 1.70 & $v_{\parallel}$  & -0.72 & -0.71 & -0.66 & $v_{x}$  & -1.88 & -2.14 & -3.91 & $v_{x}$ & 1.24-5.51i & 1.22-5.52i & 0.80-5.89i & \multirow{3}{*}{$\mathrm{eV\mathring{A}}$}\tabularnewline
$v_{\parallel}$  & 1.63 & 1.68 & 1.87 & $v_{\perp}$ & 3.08 & 3.14 & 3.34 & $v_{y}$  & 0.43 & 0.38 & 0.13 & $v_{y}$ & -0.43-0.71i & -0.42-0.68i & -0.29-0.41i & \tabularnewline
$v_{\perp}$  & 1.99 & 1.99 & 2.12 &  &  &  &  & $v_{z}$  & 1.55 & 1.66 & 2.07 & $v_{z}$ & -1.81+4.48i & -1.75+4.52i & -1.17+4.66i & \tabularnewline
\hline
$C_{\parallel}$  & 1.90 & 2.13 & 2.24 & $C_{\parallel}^{\frac{3}{2}}$  & 3.20 & 3.28 & 3.66 & $C_{x}^{p}$  & 34.44 & 31.08 & 5.91 & $C_{x}^{p}$  & 24.54 & 25.40 & 23.74 & \multirow{6}{*}{$\mathrm{eV\mathring{A}}^{2}$}\tabularnewline
$C_{\perp}$  & 5.97 & 6.26 & 8.70 & $C_{\perp}^{\frac{3}{2}}$  & -2.39 & -2.36 & -1.97 & $C_{y}^{p}$  & -9.64 & -8.90 & -5.19 & $C_{y}^{p}$  & -5.95 & -5.87 & -6.50 & \tabularnewline
$M_{\parallel}$  & 4.21 & 4.45 & 4.42 & $C_{\parallel}^{\frac{1}{2}}$  & -10.37 & -10.44 & -10.15 & $C_{z}^{p}$  & -9.52 & -8.84 & -0.12 & $C_{z}^{p}$  & 10.50 & 10.55 & 9.86 & \tabularnewline
$M_{\perp}$  & 0.99 & 0.97 & 0.60 & $C_{\perp}^{\frac{1}{2}}$  & 6.28 & 6.53 & 7.87 & $C_{x}^{s}$  & -28.54 & -35.42 & -49.83 & $C_{x}^{s}$  & -13.76 & -14.40 & -14.39 & \tabularnewline
$\gamma_{1}$  & 14.22 & 14.36 & 14.34 & $\gamma_{1}$  & 0.68 & 0.60 & 0.40 & $C_{y}^{s}$  & 4.06 & 4.08 & 3.27 & $C_{y}^{s}$  & 4.10 & 4.11 & 2.99 & \tabularnewline
$\gamma_{2}$  & 7.64 & 7.89 & 9.85 & $\gamma_{2}$  & -1.01 & -1.03 & -0.92 & $C_{z}^{s}$  & 8.82 & 7.26 & -6.94 & $C_{z}^{s}$  & -8.76 & -8.61 & -11.14 & \tabularnewline
$\         $  &      &      &      &               &       &       &            &              &      &      &       & $C_{yz}^{p}$  & 3.02 & 3.05  & 3.62   & \tabularnewline
$\         $  &      &      &      &               &       &       &            &              &      &      &       & $C_{yz}^{s}$ & -4.60 & -4.63  & -4.98 & \tabularnewline
\hline
\end{tabular}
\par\end{centering}

\protect\caption{\label{tab:4b}Parameters in the Zeeman's couplings and effective k$\cdot$p Hamiltonians for the four bands models. The parameters $g$ defining the Zeeman's couplings are dimensionless numbers, while the parameters $\Delta$, $v$, $C$ ($M$, $\gamma$) , i.e the coefficients of constants, $\mathcal{O}\left(k\right)$ terms, and $\mathcal{O}\left(k^{2}\right)$ terms, are in units of $\mathrm{eV}$, $\mathrm{eV\cdot\mathring{A}}$, and $\mathrm{eV\cdot\mathring{A}^{2}}$, respectively. Since the self-consistent LDA+mBJ potential heavily overestimates the corrections on band inversions in $\mathrm{Na_{3}Bi}$ and bismuth, here we fix the MBJ parameter as $c_{\mathrm{MBJ}}=0.93$ and $c_{\mathrm{MBJ}}=1.14$ for $\mathrm{Na_{3}Bi}$ and bismuth respectively to recover the HSE band inversion \cite{wang_Na3Bi} and experimental band gap \cite{liu_electronic_1995}.}
\end{table*}

\subsubsection{Tantalum nitride}

In the above model, the presence of quartic degenerate Dirac nodes is protected by the little group $C_{6v}$. Then a nature question is that what if the system has a lower symmetry, where, for example,the $C_{6}$ symmetry is broken? An example is the $\theta$-phase $\mathrm{TaN}$ with space group $P\bar{6}m2$, where the little group along $z$ axis is reduced to $C_{3v}$ and the Dirac nodes split into two triply degenerate nodes \cite{weng_TaN}. Specifically, the $\pm\frac{1}{2}$ states having the $C_{3}$ eigenvalues $e^{\mp i\frac{\pi}{3}}$ are still degenerate along the $z$ axis due to the presence of vertical mirrors; while the degeneracy of $\pm\frac{3}{2}$ states is no longer gauranteed since their $C_{3}$ eigenvalues are the same (-1). 
For conveniance, here we choose the basis set $\left|\frac{1}{2}\right\rangle $, $\left|\bar{\frac{1}{2}}\right\rangle $, $\left|\frac{3}{2}\right\rangle $, $\left|\bar{\frac{3}{2}}\right\rangle $ at the high symmetry point $\mathrm{A}=\left(00\pi\right)$, and get the effective k$\cdot$p model and Zeeman's coupling as
{\small{}
\begin{align}
H&^{kp}\left(\mathbf{k}\right)= \\ 
& \begin{bmatrix} \frac{\Delta}{2}+C^{\frac{1}{2}}\left(\mathbf{k}\right) & 0 \iffalse \lambda\left(\mathbf{k}\right) \fi  & \gamma_{1}k_{z}k_{+} & v_{\perp}k_{+}+\gamma_{2}k_{-}^{2}\\
* & \frac{\Delta}{2}+C^{\frac{1}{2}}\left(\mathbf{k}\right) & v_{\perp}k_{-}-\gamma_{2}k_{+}^{2} & \gamma_{1}k_{z}k_{-}\\
* & * & -\frac{\Delta}{2}+C^{\frac{3}{2}}\left(\mathbf{k}\right) & v_{\parallel}k_{z}\\
* & * & * & -\frac{\Delta}{2}+C^{\frac{3}{2}}\left(\mathbf{k}\right)
\end{bmatrix}
\end{align}
}

\begin{equation}
H^{Z}=\mu_{B}\begin{bmatrix}g_{\parallel}^{\frac{1}{2}}B_{z} & g_{\perp}^{\frac{1}{2}}B_{-} & g_{\perp}^{\prime}B_{+} & 0\\
* & -g_{\parallel}^{\frac{1}{2}}B_{z} & 0 & -g_{\perp}^{\prime}B_{-}\\
* & * & g_{\parallel}^{\frac{3}{2}}B_{z} & 0\\
* & * & * & -g_{\parallel}^{\frac{3}{2}}B_{z}
\end{bmatrix}
\end{equation}
, where the quadratic terms are defined as $C^{\frac{1}{2}/\frac{3}{2}}\left(\mathbf{k}\right)=C_{\parallel}^{\frac{1}{2}/\frac{3}{2}}k_{z}^{2}+C_{\perp}^{\frac{1}{2}/\frac{3}{2}}\left(k_{x}^{2}+k_{y}^{2}\right)$
and all the model parameters are summarized in table (\ref{tab:4b}).
Although the symmetry here is different with $\mathrm{Na_{3}Bi}$, we see that to linear order of magnetic field, the Zeeman's coupling shares the same form.

\subsubsection{ZrTe\protect\textsubscript{5}}

The transition-metal pentatelluride $\mathrm{ZrTe_{5}}$ with space group $Cmcm$ can be thought as stacked by quantum spin Hall layers with medium strength van der Waals interlayer bonding and thus is very close to the critical point between weak and strong topological phases \cite{weng_ZrTe5,Zhouj_ZrTe5}. Since the band gap is very sensitive to the external pressure and lattice constants, whether the experimental phase is in strong topology phase is still under debate \cite{ZrTe5_massless_prl,ZrTe5_2dDirac_NPG,ZrTe5_CME,ZrTe5_PT,ZrTe5_strong_prl,ZrTe5_weak_prl,ZrTe5_weak_prx}.
Probably because of this, $\mathrm{ZrTe_{5}}$ has some very nontrivial transport behaviors, such as ``anomalous'' Hall effect \cite{ZrTe5_Hall} and negative magnetoresistance \cite{ZrTe5_CME}. As discussed in Ref. {[}\onlinecite{luhz_MR_TI}{]}, a vital issue relevant to these transport properties is the interplay between Zeeman's coupling and Berry curvature. To study the impact of Zeeman's coupling, we pick up the conductance and valance bands at $\Gamma$, which form IRs $E_{\frac{1}{2}u}$ and $E_{\frac{1}{2}g}$ of the little group $D_{2h}$, as the basis to build the effective model. Take the bases order as
$\left|P\frac{1}{2}\right\rangle $, $\left|P\bar{\frac{1}{2}}\right\rangle $,
$\left|S\frac{1}{2}\right\rangle $, $\left|S\bar{\frac{1}{2}}\right\rangle $,
we get
\begin{align}
H&^{kp}\left(\mathbf{k}\right)= \\
&\begin{bmatrix}\frac{\Delta}{2}+C^{p}\left(\mathbf{k}\right) & 0 & v_{z}k_{z} & v_{x}k_{x}-iv_{y}k_{y}\\
* & \frac{\Delta}{2}+C^{p}\left(\mathbf{k}\right) & v_{x}k_{x}+iv_{y}k_{y} & -v_{z}k_{z}\\
* & * &-\frac{\Delta}{2}+C^{s}\left(\mathbf{k}\right) & 0\\
* & * & * &-\frac{\Delta}{2}+C^{s}\left(\mathbf{k}\right)
\end{bmatrix}
\end{align}
\begin{equation}
H^{Z}=\mu_{B}\begin{bmatrix}g_{z}^{p}B_{z} & g_{x}^{p}B_{x}-ig_{y}^{p}B_{y} & 0 & 0\\
* & -g_{z}^{p}B_{z} & 0 & 0\\
* & * & g_{z}^{s}B_{z} & g_{x}^{s}B_{x}-ig_{y}^{s}B_{y}\\
* & * & * & -g_{z}^{s}B_{z}
\end{bmatrix}
\end{equation}
, where the quadratic terms are defined as $C^{p/s}\left(\mathbf{k}\right)=C_{x}^{p/s}k_{x}^{2}+C_{y}^{p/s}k_{y}^{2}+C_{z}^{p/s}k_{z}^{2}$.
Different with the two models above, the presence of inversion symmetry guarantes the absence of off-block Zeeman's couplings. The first principle parameters summarized in table (\ref{tab:4b}) are calculated with the optimized crystal structure described in Ref. [\onlinecite{Zhouj_ZrTe5}].

\subsubsection{Electron pocket in bismuth}

In the above we have calculated the effective mass and g factor of the hole pocket at the T point in bismuth, to complete the discussion now let us move on to the electron pocket at L, where the conduction and valance bands form two dimensional IRs $E_{\frac{1}{2}g}$ and $E_{\frac{1}{2}u}$ of the little group $C_{2h}$ respectively due to the TR symmetry. Although the Fermi level cuts only the conduction bands, as the energy gap between conduction and valence bands is very small %
\begin{comment}
($0.011\sim0.015\mathrm{eV}$)
\end{comment}
{} and is comparable with the Fermi level energy \cite{liu_electronic_1995}, the effective model should include both of them. 
For convenience, among the three equivalent L points we pick the L point locating at the $k_{x}=0$ plane, where the $C_{2}$ axis is along the $x$ axis.
Choose the basis set as $\left|S\frac{1}{2}\right\rangle $, $\left|S\bar{\frac{1}{2}}\right\rangle $,
$\left|P\frac{1}{2}\right\rangle $, $\left|P\bar{\frac{1}{2}}\right\rangle $,
we get
\begin{align}
H&^{kp}\left(\mathbf{k}\right)= \\
&\begin{bmatrix} \frac{\Delta}{2}+C^{s}\left(\mathbf{k}\right) & 0 & v_{x}k_{x} & v_{y}k_{y}-iv_{z}k_{z}\\
* &\frac{\Delta}{2}+C^{s}\left(\mathbf{k}\right) & v_{y}^{*}k_{y}+iv_{z}^{*}k_{z} & -v_{x}^{*}k_{x}\\
* & * &-\frac{\Delta}{2}+C^{p}\left(\mathbf{k}\right) & 0\\
* & * & * &-\frac{\Delta}{2}+C^{p}\left(\mathbf{k}\right)
\end{bmatrix}
\end{align}
and
\begin{equation}
H^{Z}=\mu_{B}\begin{bmatrix}g_{x}^{s}B_{x} & g_{y}^{s}B_{y}-ig_{z}^{s}B_{z} & 0 & 0\\
* & -g_{x}^{s}B_{x} & 0 & 0\\
* & * & g_{x}^{p}B_{x} & g_{y}^{p}B_{y}-ig_{z}^{p}B_{z}\\
* & * & * & -g_{x}^{p}B_{x}
\end{bmatrix}
\end{equation}
where the quadratic terms are defined as $C^{p/s}\left(\mathbf{k}\right)=C_{x}^{p/s}k_{x}^{2}+C_{y}^{p/s}k_{y}^{2}+C_{z}^{p/s}k_{z}^{2}+C_{yz}^{p/s}k_{y}k_{z}$.
Since the $C_{2h}$ group can be got by just removing the two $C_{2}$ axes in the horizontal plane in the $D_{2h}$ group, this model shares similar form with the $\mathrm{ZrTe_{5}}$ model, except that the lower symmetry here makes all the off-diagonal parameters, i.e. $v_{x/y/z}$
and $g_{y/z}^{s/p}$, are no longer necessarily real. The calculated parameters are summarized in table (\ref{tab:4b}), where, in order to fix the gauge freedom, we have absorbed phases of $g_{y}^{s}$ and $g_{y}^{p}$ into basis and left them to be real numbers.

\section{Discussions \label{sec:discussion}}

\begin{figure}
\begin{centering}
\includegraphics[width=1\linewidth]{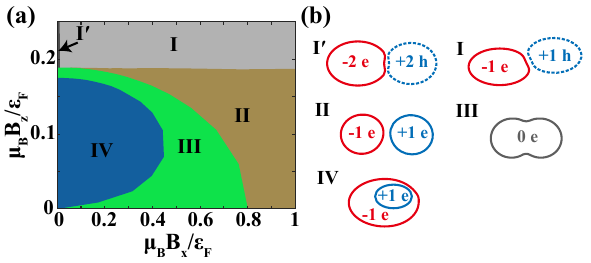}
\par\end{centering}

\protect\caption{\label{fig:PD}“Phase diagram'' of the Fermi surface topology around a single Dirac point in $\mathrm{Na_{3}Bi}$ under the Zeeman's coupling.
To get this phase diagram, the GGA parameters in table (\ref{tab:4b}) are adopted. In (a) a phase diagram consist of five phases are plotted with respect to the dimensionless parameter $\mu_{B}\mathbf{B}/\epsilon_{F}$, and in (b) diagrammatic sketches of the Fermi surface topologies for the five phases are plotted. Here the blue, grey, and red circles represent the right hand, trivial, and left hand Fermi surfaces, respectively,
where the Chern number of each Fermi surface is defined by the wavefunctions on the occupied side. The solid and dashed circles represent the electron and hole pockets, respectively. As mentioned in the text, we have presumed the clean limit and neglect $k^{2}$ terms, without which the two splitted Fermi surfaces have a nodal line crossing at the $k_{z}=0$ plane if the magnetic field is applied along $z$. However, the presence of infinite small $k^{2}$ terms will open such crossings and leave the two seperated Fermi surfaces with $\pm2$ Chern numbers, leading to the I' phase.}
\end{figure}

One of the greatest interesting points of the topological semimetals with gapless nodes or the topological insulators near critical points may be their novel responses under magnetic field due to the Berry curvarture's correction on the quasi-particle dynamics. Such responses include weak localization \cite{lu_weakloc}, negative magnetoresistance \cite{burkov_CA,burkov_NMR,son_NMR_2013,Na3Bi_NMR}, anomalous quantum oscillation \cite{luhz_SdH_2016}, and anomlous coupling with pseudo scalar phonon \cite{song_detecting_2016}, etc.
However, a related important issue, i.e. the Zeeman's coupling, which is recently revealed to play an important role in the magnetoresistance in topological insulators \cite{luhz_MR_TI}, is usually neglected in previous theoretical works. Now, with our data given above, all the magnetic response can be studied with the correction of Zeeman's coupling, in which the interplay between Berry's curvarture and Zeeman's coupling would certainly lead to interesting new physics.

Here we only show a most direct application of our method: the Fermi surface topology of Dirac semimetal under Zeeman's coupling. The intuitive picture that the Zeeman's coupling would split the double degenerate Fermi surface in Dirac semimetal into two Fermi surfaces with nonzero Chern numbers was firstly proposed to understand some magneto-transport experiments. However, no serious calculations about such splitting has been done because of the lacking of reliable g factors. 
Now we present a full evolution of the Fermi surface around a single Dirac point in $\mathrm{Na_{3}Bi}$ with respect to the strength and direction of the magnetic field. Since there are only two relevant energy scales, i.e. the Fermi energy $\epsilon_{F}$ and the Zeeman's splitting $\mu_{B}\left|\mathbf{B}\right|$, if we presume the clean limit and keep only $k$ linear terms in Eq.(\ref{eq:Hk-Na3Bi}), the model can be rescaled such that the Fermi surface topology will only depend on the dimensionless parameter $\mu_{B}\mathbf{B}/\epsilon_{F}$.
The result is summarized as a ``phase diagram'' shown in Fig. (\ref{fig:PD}), where different ``phases'' are distinguished by the carrior types (electron or hole) and Chern numbers of the Fermi surfaces, and the magnetic field is applied only within the $xz$ plane since all the vertical planes are equivalent with each other. 
With different magnetic field, various of phases, including the phase with one trivial electron pocket, the phase with two seperate nontrivial electron pockets, the phase with two nested nontrivial electron pockets, and the phase with a nontrivial electron pocket plus a nontrivial hole pocket, can be achieved, leading to a fruitful phase diagram.

\section{Summary\label{sec:summary}}
In summary, the first principle method for the calculation of the effective g-factor tensor in solids has been proposed for the first time, and the coupling parameters in a few topological materials have been calculated and discussed.
In the framework of quasi-degenerate perturbation theory, we define the effective Zeeman's coupling in magnetic field as the gauge invariant part of the second order perturbation Hamiltonian and express all the parameters in it by the momentum and spin matrix elements among the Bloch states, which can be computed within the PAW formulation of the DFT in this work.
To verify the validity of this method, the Landé g-factors of the two bands systems, i.e. the hole pocket in bismuth and the conduction bands in Bi$_2$Se$_3$, whose Landé  g-factors have been measured in experiments, are calculated and results comparable with experimental data are obtained.
Furthermore, we also derive, discuss, and calculate the effective Zeeman's couplings for a few multi-bands models in topological materials such as Na$_3$Bi, TaN, and ZrTe$_5$.
As a simple application of our data, a full evolution of the Fermi surface topology in Dirac semimetal under Zeeman's coupling with respect to the strength and direction of magnetic field is obtained and discussed based on the data of Na$_3$Bi.
Further theoretical investigations for other applications, such as the interplay of the Zeeman's coupling and the Berry phase in magnetotransport experiments, and the role of multi-bands Zeeman's coupling in quantum oscillations are strongly encouraged.
%In the end, we give a brief summary in section \ref{sec:summary}.

\section{Acknowledgement}

This work is supported by the National Natural Science Foundation of China, the National 973 program of China (Grant No. 2013CB921700) and the “Strategic Priority Research Program (B)” of the Chinese Academy
of Sciences (Grant No. XDB07020100).

\appendix

\section{PAW formulation of the matrix elements\label{sec:PAW}}

In the PAW formulation adopt in VASP, the matrix element of operator
$\hat{O}$ can be calculated as \cite{blochl_PAW,kresse_PAW}
\begin{align}
O_{nn^{\prime}} & =\langle\psi_{n\mathbf{K}}|\hat{O}|\psi_{n^{\prime}\mathbf{K}}\rangle\nonumber \\
 & =\langle\widetilde{\psi}_{n\mathbf{K}}|\hat{O}|\widetilde{\psi}_{n^{\prime}\mathbf{K}}\rangle\nonumber \\
 & +\sum_{a\mu\nu}\sum_{\zeta\zeta^{\prime}}\langle\widetilde{\psi}_{n\mathbf{K}}|\widetilde{p}_{a\mu}\zeta\big\rangle O_{\mu\zeta,\nu\zeta^{\prime}}^{a}\big\langle\widetilde{p}_{a\nu}\zeta^{\prime}|\widetilde{\psi}_{n^{\prime}\mathbf{K}}\rangle\label{eq:O}
\end{align}
, where $|\widetilde{\psi}_{n\mathbf{K}}\rangle$ is the pseudo Bloch
wavefunction, $|\widetilde{p}_{a\mu}\zeta\rangle$ is a projector
wavefunction at the $a$-th atom consisting of a real space projector
wavefunction $\tilde{p}_{a\mu}\left(\mathbf{r}\right)$ and a spinor
wavefunction $\left|\zeta\right\rangle $ ($\zeta=\uparrow\downarrow$),
and $O_{\mu\nu}^{a}$ is the PAW matrix of $\hat{O}$ in the $a$-th
atom's augmentation sphere. Here the pseudo Bloch wavefunction is
spanned by the plane waves
\begin{equation}
\psi_{n\mathbf{K}}\left(\mathbf{r}\zeta\right)=\sum_{\mathbf{G}}c_{\zeta,\mathbf{G}}^{n\mathbf{K}}e^{i\left(\mathbf{K}+\mathbf{G}\right)\cdot\mathbf{r}}
\end{equation}
, and the PAW matrix of $\hat{O}$ is defined as
\begin{equation}
O_{\mu\zeta,\nu\zeta^{\prime}}^{a}=\langle\phi_{a\mu}\zeta|\hat{O}|\phi_{a\nu}\zeta^{\prime}\rangle-\langle\widetilde{\phi}_{a\mu}\zeta|\hat{O}|\widetilde{\phi}_{a\nu}\zeta^{\prime}\rangle
\end{equation}
, with $\phi_{a\mu}\left(\mathbf{r}\right)$ the all electron partial
wavefunction, and $\widetilde{\phi}_{a\mu}\left(\mathbf{r}\right)$
the pseudo partial wavefunction, both of which are stored as an angular
part and a radial part
\begin{equation}
\phi_{a\mu}\left(\mathbf{r}\right)=Y_{l_{\mu}}^{m_{\mu}}\left(\widehat{\mathbf{r}-\mathbf{R}_{a}}\right)R_{a\mu}\left(\left|\mathbf{r}-\mathbf{R}_{a}\right|\right)
\end{equation}
\begin{equation}
\widetilde{\phi}_{a\mu}\left(\mathbf{r}\right)=Y_{l_{\mu}}^{m_{\mu}}\left(\widehat{\mathbf{r}-\mathbf{R}_{a}}\right)\widetilde{R}_{a\mu}\left(\left|\mathbf{r}-\mathbf{R}_{a}\right|\right)
\end{equation}
, where $\mathbf{R}_{a}$ is the location of the $a$-th atom. As
both the plane wave coefficients $c_{\zeta,\mathbf{G}}^{n\mathbf{K}}$
and the projection coefficients $\langle\widetilde{p}_{a\mu}\zeta|\widetilde{\psi}_{n\mathbf{K}}\big\rangle$
have been calculated in the VASP code, to calculate the matrix element
of $\hat{O}$ we only need to calculate the pseudo wavefunction contribution,
i.e. the first term in Eq. (\ref{eq:O}), and the PAW matrix elements
separately and then substitute them back to Eq. (\ref{eq:O}).

The spin elements are very direct to calculate. The pseudo wavefunction
contribution is
\begin{equation}
\langle\widetilde{\psi}_{n\mathbf{K}}|\hat{\mathbf{s}}|\widetilde{\psi}_{n^{\prime}\mathbf{K}}\rangle=\frac{\hbar}{2}\sum_{\mathbf{G}\zeta\zeta^{\prime}}\boldsymbol{\sigma}_{\zeta,\zeta^{\prime}}c_{\zeta,\mathbf{G}}^{n\mathbf{K}*}c_{\zeta^{\prime},\mathbf{G}}^{n^{\prime}\mathbf{K}}
\end{equation}
, and the PAW matrix is
\begin{align}
\mathbf{s}_{\mu\zeta,\nu\zeta^{\prime}}^{a} & =\delta_{l_{\mu}l_{\nu}}\delta_{m_{\mu}m_{\nu}}\frac{\hbar}{2}\sum_{\zeta\zeta^{\prime}}\boldsymbol{\sigma}_{\zeta,\zeta^{\prime}}\int dr\cdot r^{2}R_{a\mu}^{*}R_{a\nu}\nonumber \\
 & -\text{the integral with }\widetilde{R}
\end{align}
, where the integral over $r$ is evaluated on a logarithmic radial
grid. Nevertheless, the calculation of the momentum element is a bit
more complicated. According to Eq. (\ref{eq:O}), the momentum element
can be directly written as
\begin{widetext}
\begin{align}
\boldsymbol{\pi}_{nn^{\prime}} & =\langle\widetilde{\psi}_{n\mathbf{K}}|\hat{\mathbf{p}}|\widetilde{\psi}_{n^{\prime}\mathbf{K}}\rangle+\frac{1}{2m_{e}e^{2}}\langle\widetilde{\psi}_{n\mathbf{K}}|\hat{\mathbf{s}}\times\nabla V|\widetilde{\psi}_{n^{\prime}\mathbf{K}}\rangle\nonumber \\
 & +\sum_{a\mu\nu}\sum_{\zeta\zeta^{\prime}}\langle\widetilde{\psi}_{n\mathbf{K}}|\widetilde{p}_{a\mu}\zeta\big\rangle\left[\langle\phi_{a\mu}|\hat{\mathbf{p}}|\phi_{a\nu}\rangle-\langle\widetilde{\phi}_{a\mu}|\hat{\mathbf{p}}|\widetilde{\phi}_{a\nu}\rangle\right]\big\langle\widetilde{p}_{a\nu}\zeta^{\prime}|\widetilde{\psi}_{n^{\prime}\mathbf{K}}\rangle\nonumber \\
 & +\frac{1}{2m_{e}e^{2}}\sum_{a\mu\nu}\sum_{\zeta\zeta^{\prime}}\langle\widetilde{\psi}_{n\mathbf{K}}|\widetilde{p}_{a\mu}\zeta\big\rangle\left[\langle\phi_{a\mu}\zeta|\hat{\mathbf{s}}\times\nabla V|\zeta^{\prime}\phi_{a\nu}\rangle-\langle\widetilde{\phi}_{a\mu}\zeta|\hat{\mathbf{s}}\times\nabla V|\widetilde{\phi}_{a\nu}\zeta^{\prime}\rangle\right]\big\langle\widetilde{p}_{a\nu}\zeta^{\prime}|\widetilde{\psi}_{n^{\prime}\mathbf{K}}\rangle
\end{align}

\end{widetext}

\noindent It should be noticed that the second term in the first line
and the second term in the third line will cancel each other because
in VASP the SOC effect is considered only within the augmentation
spheres, in which the pseudo Bloch wavefunctions equal to their projections
on pseudo partial wavefunctions, i.e. $|\widetilde{\psi}_{n\mathbf{K}}\rangle=\sum_{a\mu\zeta}|\widetilde{\phi}_{a\mu}\zeta\rangle\big\langle\widetilde{p}_{a\nu}\zeta|\widetilde{\psi}_{n\mathbf{K}}\rangle$.
Therefore, the momentum element is simplified to
\begin{align}
\boldsymbol{\pi}_{nn^{\prime}} & =\langle\widetilde{\psi}_{n\mathbf{K}}|\hat{\mathbf{p}}|\widetilde{\psi}_{n^{\prime}\mathbf{K}}\rangle\nonumber \\
 & +\sum_{a\mu\nu}\sum_{\zeta\zeta^{\prime}}\langle\widetilde{\psi}_{n\mathbf{K}}|\widetilde{p}_{a\mu}\zeta\big\rangle\boldsymbol{\pi}_{\mu\zeta,\nu\zeta^{\prime}}^{a\prime}\big\langle\widetilde{p}_{a\nu}\zeta^{\prime}|\widetilde{\psi}_{n^{\prime}\mathbf{K}}\rangle
\end{align}
, where the psuedo wavefunction contribution is
\begin{align}
\langle\widetilde{\psi}_{n\mathbf{K}}|\hat{\mathbf{p}}|\widetilde{\psi}_{n^{\prime}\mathbf{K}}\rangle & =\sum_{\mathbf{G}\zeta}\hbar\left(\mathbf{K}+\mathbf{G}\right)c_{\zeta,\mathbf{G}}^{n\mathbf{K}*}c_{\zeta,\mathbf{G}}^{n^{\prime}\mathbf{K}}
\end{align}
, and $\boldsymbol{\pi}_{\mu\zeta,\nu\zeta^{\prime}}^{a\prime}$ is
defined as
\begin{align}
\boldsymbol{\pi}_{\mu\zeta,\nu\zeta^{\prime}}^{a\prime} & =\delta_{\zeta\zeta^{\prime}}\langle\phi_{a\mu}|\hat{\mathbf{p}}|\phi_{a\mu}\rangle-\delta_{\zeta\zeta^{\prime}}\langle\widetilde{\phi}_{a\mu}|\hat{p}^{i}|\widetilde{\phi}_{a\mu}\rangle\nonumber \\
 & +\frac{\hbar}{2m_{e}e^{2}}\cdot\frac{\hbar}{2}\boldsymbol{\sigma}_{\zeta,\zeta^{\prime}}\times\langle\phi_{a\mu}|\nabla V|\phi_{a\mu}\rangle
\end{align}
All the integrals between partial wavefunctions can be decomposed
as an angular part and a radial part, such as
\begin{align}
\langle\phi_{a\mu}|\hat{\mathbf{p}}|\phi_{a\mu}\rangle & =-i\hbar\int d\Omega Y_{l_{\mu}}^{m_{\mu}*}\nabla Y_{l_{\nu}}^{m_{\nu}}\int dr\cdot r^{2}R_{a\mu}^{*}R_{a\nu}\nonumber \\
 & -i\hbar\int d\Omega Y_{l_{\mu}}^{m_{\mu}*}\frac{\mathbf{r}}{r}Y_{l_{\nu}}^{m_{\nu}}\int dr\cdot r^{2}R_{a\mu}^{*}\partial_{r}R_{a\nu}
\end{align}
and
\begin{equation}
\langle\phi_{a\mu}|\nabla V|\phi_{a\mu}\rangle\approx\int d\Omega Y_{l_{\mu}}^{m_{\mu}*}\frac{\mathbf{r}}{r}Y_{l_{\nu}}^{m_{\nu}}\int dr\cdot r^{2}R_{a\mu}^{*}\partial_{r}VR_{a\nu}
\end{equation}
In practice, the angular integrals concerning with spherical harmonics
are evaluated with the help of the code in asa.F of VASP.

\bibliographystyle{apsrev4-1}
\bibliography{main}

\begin{thebibliography}{46}%
\makeatletter
\providecommand \@ifxundefined [1]{%
 \@ifx{#1\undefined}
}%
\providecommand \@ifnum [1]{%
 \ifnum #1\expandafter \@firstoftwo
 \else \expandafter \@secondoftwo
 \fi
}%
\providecommand \@ifx [1]{%
 \ifx #1\expandafter \@firstoftwo
 \else \expandafter \@secondoftwo
 \fi
}%
\providecommand \natexlab [1]{#1}%
\providecommand \enquote  [1]{``#1''}%
\providecommand \bibnamefont  [1]{#1}%
\providecommand \bibfnamefont [1]{#1}%
\providecommand \citenamefont [1]{#1}%
\providecommand \href@noop [0]{\@secondoftwo}%
\providecommand \href [0]{\begingroup \@sanitize@url \@href}%
\providecommand \@href[1]{\@@startlink{#1}\@@href}%
\providecommand \@@href[1]{\endgroup#1\@@endlink}%
\providecommand \@sanitize@url [0]{\catcode `\\12\catcode `\$12\catcode
  `\&12\catcode `\#12\catcode `\^12\catcode `\_12\catcode `\%12\relax}%
\providecommand \@@startlink[1]{}%
\providecommand \@@endlink[0]{}%
\providecommand \url  [0]{\begingroup\@sanitize@url \@url }%
\providecommand \@url [1]{\endgroup\@href {#1}{\urlprefix }}%
\providecommand \urlprefix  [0]{URL }%
\providecommand \Eprint [0]{\href }%
\providecommand \doibase [0]{http://dx.doi.org/}%
\providecommand \selectlanguage [0]{\@gobble}%
\providecommand \bibinfo  [0]{\@secondoftwo}%
\providecommand \bibfield  [0]{\@secondoftwo}%
\providecommand \translation [1]{[#1]}%
\providecommand \BibitemOpen [0]{}%
\providecommand \bibitemStop [0]{}%
\providecommand \bibitemNoStop [0]{.\EOS\space}%
\providecommand \EOS [0]{\spacefactor3000\relax}%
\providecommand \BibitemShut  [1]{\csname bibitem#1\endcsname}%
\let\auto@bib@innerbib\@empty
%</preamble>
\bibitem [{\citenamefont {Ashcroft}\ and\ \citenamefont
  {Mermin}(1976)}]{ashcroft_solid_book}%
  \BibitemOpen
  \bibfield  {author} {\bibinfo {author} {\bibfnamefont {N.~W.}\ \bibnamefont
  {Ashcroft}}\ and\ \bibinfo {author} {\bibfnamefont {N.~D.}\ \bibnamefont
  {Mermin}},\ }\href@noop {} {\emph {\bibinfo {title} {Solid {State}
  {Physics}}}}\ (\bibinfo  {publisher} {Holt, Rinehart and Winston},\ \bibinfo
  {year} {1976})\BibitemShut {NoStop}%
\bibitem [{\citenamefont {Shoenberg}(2009)}]{oscillation_book}%
  \BibitemOpen
  \bibfield  {author} {\bibinfo {author} {\bibfnamefont {D.}~\bibnamefont
  {Shoenberg}},\ }\href@noop {} {\emph {\bibinfo {title} {Magnetic
  {Oscillations} in {Metals}}}}\ (\bibinfo {year} {2009})\BibitemShut {NoStop}%
\bibitem [{\citenamefont {Luttinger}(1956)}]{luttinger_quantum_1956}%
  \BibitemOpen
  \bibfield  {author} {\bibinfo {author} {\bibfnamefont {J.~M.}\ \bibnamefont
  {Luttinger}},\ }\href {\doibase 10.1103/PhysRev.102.1030} {\bibfield
  {journal} {\bibinfo  {journal} {Physical Review}\ }\textbf {\bibinfo {volume}
  {102}},\ \bibinfo {pages} {1030} (\bibinfo {year} {1956})}\BibitemShut
  {NoStop}%
\bibitem [{\citenamefont {Cohen}\ and\ \citenamefont
  {Blount}(1960)}]{cohen_g-factor_1960}%
  \BibitemOpen
  \bibfield  {author} {\bibinfo {author} {\bibfnamefont {M.~H.}\ \bibnamefont
  {Cohen}}\ and\ \bibinfo {author} {\bibfnamefont {E.~I.}\ \bibnamefont
  {Blount}},\ }\href {\doibase 10.1080/14786436008243294} {\bibfield  {journal}
  {\bibinfo  {journal} {Philosophical Magazine}\ }\textbf {\bibinfo {volume}
  {5}},\ \bibinfo {pages} {115} (\bibinfo {year} {1960})}\BibitemShut {NoStop}%
\bibitem [{\citenamefont {Chang}\ and\ \citenamefont
  {Niu}(1996)}]{chang_berry_1996}%
  \BibitemOpen
  \bibfield  {author} {\bibinfo {author} {\bibfnamefont {M.-C.}\ \bibnamefont
  {Chang}}\ and\ \bibinfo {author} {\bibfnamefont {Q.}~\bibnamefont {Niu}},\
  }\href {\doibase 10.1103/PhysRevB.53.7010} {\bibfield  {journal} {\bibinfo
  {journal} {Physical Review B}\ }\textbf {\bibinfo {volume} {53}},\ \bibinfo
  {pages} {7010} (\bibinfo {year} {1996})}\BibitemShut {NoStop}%
\bibitem [{\citenamefont {Chang}\ and\ \citenamefont
  {Niu}(2008)}]{chang_berry_2008}%
  \BibitemOpen
  \bibfield  {author} {\bibinfo {author} {\bibfnamefont {M.-C.}\ \bibnamefont
  {Chang}}\ and\ \bibinfo {author} {\bibfnamefont {Q.}~\bibnamefont {Niu}},\
  }\href {\doibase 10.1088/0953-8984/20/19/193202} {\bibfield  {journal}
  {\bibinfo  {journal} {Journal of Physics: Condensed Matter}\ }\textbf
  {\bibinfo {volume} {20}},\ \bibinfo {pages} {193202} (\bibinfo {year}
  {2008})}\BibitemShut {NoStop}%
\bibitem [{\citenamefont {Lu}\ and\ \citenamefont {Shen}(2015)}]{lu_weakloc}%
  \BibitemOpen
  \bibfield  {author} {\bibinfo {author} {\bibfnamefont {H.-Z.}\ \bibnamefont
  {Lu}}\ and\ \bibinfo {author} {\bibfnamefont {S.-Q.}\ \bibnamefont {Shen}},\
  }\href {\doibase 10.1103/PhysRevB.92.035203} {\bibfield  {journal} {\bibinfo
  {journal} {Physical Review B}\ }\textbf {\bibinfo {volume} {92}} (\bibinfo
  {year} {2015}),\ 10.1103/PhysRevB.92.035203}\BibitemShut {NoStop}%
\bibitem [{\citenamefont {Nielsen}\ and\ \citenamefont
  {Ninomiya}(1983)}]{abj_1983}%
  \BibitemOpen
  \bibfield  {author} {\bibinfo {author} {\bibfnamefont {H.~B.}\ \bibnamefont
  {Nielsen}}\ and\ \bibinfo {author} {\bibfnamefont {M.}~\bibnamefont
  {Ninomiya}},\ }\href {\doibase 10.1016/0370-2693(83)91529-0} {\bibfield
  {journal} {\bibinfo  {journal} {Physics Letters B}\ }\textbf {\bibinfo
  {volume} {130}},\ \bibinfo {pages} {389} (\bibinfo {year}
  {1983})}\BibitemShut {NoStop}%
\bibitem [{\citenamefont {Burkov}(2014)}]{burkov_CA}%
  \BibitemOpen
  \bibfield  {author} {\bibinfo {author} {\bibfnamefont {A.}~\bibnamefont
  {Burkov}},\ }\href {\doibase 10.1103/PhysRevLett.113.247203} {\bibfield
  {journal} {\bibinfo  {journal} {Physical Review Letters}\ }\textbf {\bibinfo
  {volume} {113}},\ \bibinfo {pages} {247203} (\bibinfo {year}
  {2014})}\BibitemShut {NoStop}%
\bibitem [{\citenamefont {Burkov}(2015)}]{burkov_NMR}%
  \BibitemOpen
  \bibfield  {author} {\bibinfo {author} {\bibfnamefont {A.~A.}\ \bibnamefont
  {Burkov}},\ }\href {\doibase 10.1103/PhysRevB.91.245157} {\bibfield
  {journal} {\bibinfo  {journal} {Physical Review B}\ }\textbf {\bibinfo
  {volume} {91}},\ \bibinfo {pages} {245157} (\bibinfo {year}
  {2015})}\BibitemShut {NoStop}%
\bibitem [{\citenamefont {Son}\ and\ \citenamefont
  {Spivak}(2013)}]{son_NMR_2013}%
  \BibitemOpen
  \bibfield  {author} {\bibinfo {author} {\bibfnamefont {D.~T.}\ \bibnamefont
  {Son}}\ and\ \bibinfo {author} {\bibfnamefont {B.~Z.}\ \bibnamefont
  {Spivak}},\ }\href {\doibase 10.1103/PhysRevB.88.104412} {\bibfield
  {journal} {\bibinfo  {journal} {Physical Review B}\ }\textbf {\bibinfo
  {volume} {88}},\ \bibinfo {pages} {104412} (\bibinfo {year}
  {2013})}\BibitemShut {NoStop}%
\bibitem [{\citenamefont {Xiong}\ \emph {et~al.}(2015)\citenamefont {Xiong},
  \citenamefont {Kushwaha}, \citenamefont {Liang}, \citenamefont {Krizan},
  \citenamefont {Hirschberger}, \citenamefont {Wang}, \citenamefont {Cava},\
  and\ \citenamefont {Ong}}]{Na3Bi_NMR}%
  \BibitemOpen
  \bibfield  {author} {\bibinfo {author} {\bibfnamefont {J.}~\bibnamefont
  {Xiong}}, \bibinfo {author} {\bibfnamefont {S.~K.}\ \bibnamefont {Kushwaha}},
  \bibinfo {author} {\bibfnamefont {T.}~\bibnamefont {Liang}}, \bibinfo
  {author} {\bibfnamefont {J.~W.}\ \bibnamefont {Krizan}}, \bibinfo {author}
  {\bibfnamefont {M.}~\bibnamefont {Hirschberger}}, \bibinfo {author}
  {\bibfnamefont {W.}~\bibnamefont {Wang}}, \bibinfo {author} {\bibfnamefont
  {R.~J.}\ \bibnamefont {Cava}}, \ and\ \bibinfo {author} {\bibfnamefont
  {N.~P.}\ \bibnamefont {Ong}},\ }\href {\doibase 10.1126/science.aac6089}
  {\bibfield  {journal} {\bibinfo  {journal} {Science}\ }\textbf {\bibinfo
  {volume} {350}},\ \bibinfo {pages} {413} (\bibinfo {year}
  {2015})}\BibitemShut {NoStop}%
\bibitem [{\citenamefont {Parameswaran}\ \emph {et~al.}(2014)\citenamefont
  {Parameswaran}, \citenamefont {Grover}, \citenamefont {Abanin}, \citenamefont
  {Pesin},\ and\ \citenamefont {Vishwanath}}]{CA_nonlocal}%
  \BibitemOpen
  \bibfield  {author} {\bibinfo {author} {\bibfnamefont {S.}~\bibnamefont
  {Parameswaran}}, \bibinfo {author} {\bibfnamefont {T.}~\bibnamefont
  {Grover}}, \bibinfo {author} {\bibfnamefont {D.}~\bibnamefont {Abanin}},
  \bibinfo {author} {\bibfnamefont {D.}~\bibnamefont {Pesin}}, \ and\ \bibinfo
  {author} {\bibfnamefont {A.}~\bibnamefont {Vishwanath}},\ }\href {\doibase
  10.1103/PhysRevX.4.031035} {\bibfield  {journal} {\bibinfo  {journal}
  {Physical Review X}\ }\textbf {\bibinfo {volume} {4}},\ \bibinfo {pages}
  {031035} (\bibinfo {year} {2014})}\BibitemShut {NoStop}%
\bibitem [{\citenamefont {Wang}\ \emph {et~al.}(2016)\citenamefont {Wang},
  \citenamefont {Lu},\ and\ \citenamefont {Shen}}]{luhz_SdH_2016}%
  \BibitemOpen
  \bibfield  {author} {\bibinfo {author} {\bibfnamefont {C.}~\bibnamefont
  {Wang}}, \bibinfo {author} {\bibfnamefont {H.-Z.}\ \bibnamefont {Lu}}, \ and\
  \bibinfo {author} {\bibfnamefont {S.-Q.}\ \bibnamefont {Shen}},\ }\href
  {\doibase 10.1103/PhysRevLett.117.077201} {\bibfield  {journal} {\bibinfo
  {journal} {Physical Review Letters}\ }\textbf {\bibinfo {volume} {117}},\
  \bibinfo {pages} {077201} (\bibinfo {year} {2016})}\BibitemShut {NoStop}%
\bibitem [{\citenamefont {Song}\ \emph {et~al.}(2016)\citenamefont {Song},
  \citenamefont {Zhao}, \citenamefont {Fang},\ and\ \citenamefont
  {Dai}}]{song_detecting_2016}%
  \BibitemOpen
  \bibfield  {author} {\bibinfo {author} {\bibfnamefont {Z.}~\bibnamefont
  {Song}}, \bibinfo {author} {\bibfnamefont {J.}~\bibnamefont {Zhao}}, \bibinfo
  {author} {\bibfnamefont {Z.}~\bibnamefont {Fang}}, \ and\ \bibinfo {author}
  {\bibfnamefont {X.}~\bibnamefont {Dai}},\ }\href {\doibase
  10.1103/PhysRevB.94.214306} {\bibfield  {journal} {\bibinfo  {journal} {Phys.
  Rev. B}\ }\textbf {\bibinfo {volume} {94}},\ \bibinfo {pages} {214306}
  (\bibinfo {year} {2016})}\BibitemShut {NoStop}%
\bibitem [{\citenamefont {Dai}\ \emph {et~al.}(2017)\citenamefont {Dai},
  \citenamefont {Du},\ and\ \citenamefont {Lu}}]{luhz_MR_TI}%
  \BibitemOpen
  \bibfield  {author} {\bibinfo {author} {\bibfnamefont {X.}~\bibnamefont
  {Dai}}, \bibinfo {author} {\bibfnamefont {Z.~Z.}\ \bibnamefont {Du}}, \ and\
  \bibinfo {author} {\bibfnamefont {H.-Z.}\ \bibnamefont {Lu}},\ }\href
  {http://arxiv.org/abs/1705.02724} {\bibfield  {journal} {\bibinfo  {journal}
  {arXiv:1705.02724 [cond-mat]}\ } (\bibinfo {year} {2017})},\ \bibinfo {note}
  {arXiv: 1705.02724}\BibitemShut {NoStop}%
\bibitem [{\citenamefont {L{\"o}wdin}(1951)}]{lowdin1951note}%
  \BibitemOpen
  \bibfield  {author} {\bibinfo {author} {\bibfnamefont {P.-O.}\ \bibnamefont
  {L{\"o}wdin}},\ }\href {http://dx.doi.org/10.1063/1.1748067} {\bibfield
  {journal} {\bibinfo  {journal} {The Journal of Chemical Physics}\ }\textbf
  {\bibinfo {volume} {19}},\ \bibinfo {pages} {1396} (\bibinfo {year}
  {1951})}\BibitemShut {NoStop}%
\bibitem [{\citenamefont {Luttinger}\ and\ \citenamefont
  {Kohn}(1955)}]{luttinger_perturbation}%
  \BibitemOpen
  \bibfield  {author} {\bibinfo {author} {\bibfnamefont {J.~M.}\ \bibnamefont
  {Luttinger}}\ and\ \bibinfo {author} {\bibfnamefont {W.}~\bibnamefont
  {Kohn}},\ }\href {\doibase 10.1103/PhysRev.97.869} {\bibfield  {journal}
  {\bibinfo  {journal} {Physical Review}\ }\textbf {\bibinfo {volume} {97}},\
  \bibinfo {pages} {869} (\bibinfo {year} {1955})}\BibitemShut {NoStop}%
\bibitem [{\citenamefont {Winkler}(2003)}]{winkler_spin-orbit_2003}%
  \BibitemOpen
  \bibfield  {author} {\bibinfo {author} {\bibfnamefont {R.}~\bibnamefont
  {Winkler}},\ }\href@noop {} {\emph {\bibinfo {title} {Spin-orbit {Coupling}
  {Effects} in {Two}-{Dimensional} {Electron} and {Hole} {Systems}}}}\
  (\bibinfo  {publisher} {Springer Science \& Business Media},\ \bibinfo {year}
  {2003})\BibitemShut {NoStop}%
\bibitem [{\citenamefont {Fuseya}\ \emph {et~al.}(2015)\citenamefont {Fuseya},
  \citenamefont {Zhu}, \citenamefont {Fauqué}, \citenamefont {Kang},
  \citenamefont {Lenoir},\ and\ \citenamefont {Behnia}}]{Yuki_gfactor}%
  \BibitemOpen
  \bibfield  {author} {\bibinfo {author} {\bibfnamefont {Y.}~\bibnamefont
  {Fuseya}}, \bibinfo {author} {\bibfnamefont {Z.}~\bibnamefont {Zhu}},
  \bibinfo {author} {\bibfnamefont {B.}~\bibnamefont {Fauqué}}, \bibinfo
  {author} {\bibfnamefont {W.}~\bibnamefont {Kang}}, \bibinfo {author}
  {\bibfnamefont {B.}~\bibnamefont {Lenoir}}, \ and\ \bibinfo {author}
  {\bibfnamefont {K.}~\bibnamefont {Behnia}},\ }\href {\doibase
  10.1103/PhysRevLett.115.216401} {\bibfield  {journal} {\bibinfo  {journal}
  {Physical Review Letters}\ }\textbf {\bibinfo {volume} {115}},\ \bibinfo
  {pages} {216401} (\bibinfo {year} {2015})}\BibitemShut {NoStop}%
\bibitem [{\citenamefont {Kohn}\ and\ \citenamefont {Sham}(1965)}]{Kohn_Sham}%
  \BibitemOpen
  \bibfield  {author} {\bibinfo {author} {\bibfnamefont {W.}~\bibnamefont
  {Kohn}}\ and\ \bibinfo {author} {\bibfnamefont {L.~J.}\ \bibnamefont
  {Sham}},\ }\href {\doibase 10.1103/PhysRev.140.A1133} {\bibfield  {journal}
  {\bibinfo  {journal} {Phys. Rev.}\ }\textbf {\bibinfo {volume} {140}},\
  \bibinfo {pages} {A1133} (\bibinfo {year} {1965})}\BibitemShut {NoStop}%
\bibitem [{\citenamefont {Ceperley}\ and\ \citenamefont
  {Alder}(1980)}]{LDA_CA}%
  \BibitemOpen
  \bibfield  {author} {\bibinfo {author} {\bibfnamefont {D.~M.}\ \bibnamefont
  {Ceperley}}\ and\ \bibinfo {author} {\bibfnamefont {B.~J.}\ \bibnamefont
  {Alder}},\ }\href {\doibase 10.1103/PhysRevLett.45.566} {\bibfield  {journal}
  {\bibinfo  {journal} {Phys. Rev. Lett.}\ }\textbf {\bibinfo {volume} {45}},\
  \bibinfo {pages} {566} (\bibinfo {year} {1980})}\BibitemShut {NoStop}%
\bibitem [{\citenamefont {Perdew}\ \emph {et~al.}(1996)\citenamefont {Perdew},
  \citenamefont {Burke},\ and\ \citenamefont {Ernzerhof}}]{perdew_GGA}%
  \BibitemOpen
  \bibfield  {author} {\bibinfo {author} {\bibfnamefont {J.~P.}\ \bibnamefont
  {Perdew}}, \bibinfo {author} {\bibfnamefont {K.}~\bibnamefont {Burke}}, \
  and\ \bibinfo {author} {\bibfnamefont {M.}~\bibnamefont {Ernzerhof}},\ }\href
  {\doibase 10.1103/PhysRevLett.77.3865} {\bibfield  {journal} {\bibinfo
  {journal} {Physical Review Letters}\ }\textbf {\bibinfo {volume} {77}},\
  \bibinfo {pages} {3865} (\bibinfo {year} {1996})}\BibitemShut {NoStop}%
\bibitem [{\citenamefont {Tran}\ and\ \citenamefont {Blaha}(2009)}]{mBJ_2009}%
  \BibitemOpen
  \bibfield  {author} {\bibinfo {author} {\bibfnamefont {F.}~\bibnamefont
  {Tran}}\ and\ \bibinfo {author} {\bibfnamefont {P.}~\bibnamefont {Blaha}},\
  }\href {\doibase 10.1103/PhysRevLett.102.226401} {\bibfield  {journal}
  {\bibinfo  {journal} {Physical Review Letters}\ }\textbf {\bibinfo {volume}
  {102}},\ \bibinfo {pages} {226401} (\bibinfo {year} {2009})}\BibitemShut
  {NoStop}%
\bibitem [{\citenamefont {Verdún}\ and\ \citenamefont
  {Drew}(1976)}]{verdun_far-infrared_1976}%
  \BibitemOpen
  \bibfield  {author} {\bibinfo {author} {\bibfnamefont {H.~R.}\ \bibnamefont
  {Verdún}}\ and\ \bibinfo {author} {\bibfnamefont {H.~D.}\ \bibnamefont
  {Drew}},\ }\href {\doibase 10.1103/PhysRevB.14.1370} {\bibfield  {journal}
  {\bibinfo  {journal} {Physical Review B}\ }\textbf {\bibinfo {volume} {14}},\
  \bibinfo {pages} {1370} (\bibinfo {year} {1976})}\BibitemShut {NoStop}%
\bibitem [{\citenamefont {Köhler}\ and\ \citenamefont
  {Wöchner}(1975)}]{gfactor_Bi2Se3}%
  \BibitemOpen
  \bibfield  {author} {\bibinfo {author} {\bibfnamefont {H.}~\bibnamefont
  {Köhler}}\ and\ \bibinfo {author} {\bibfnamefont {E.}~\bibnamefont
  {Wöchner}},\ }\href {\doibase 10.1002/pssb.2220670229} {\bibfield  {journal}
  {\bibinfo  {journal} {physica status solidi (b)}\ }\textbf {\bibinfo {volume}
  {67}},\ \bibinfo {pages} {665} (\bibinfo {year} {1975})}\BibitemShut
  {NoStop}%
\bibitem [{\citenamefont {Alicea}\ and\ \citenamefont
  {Balents}(2009)}]{alicea_bismuth_2009}%
  \BibitemOpen
  \bibfield  {author} {\bibinfo {author} {\bibfnamefont {J.}~\bibnamefont
  {Alicea}}\ and\ \bibinfo {author} {\bibfnamefont {L.}~\bibnamefont
  {Balents}},\ }\href {\doibase 10.1103/PhysRevB.79.241101} {\bibfield
  {journal} {\bibinfo  {journal} {Physical Review B}\ }\textbf {\bibinfo
  {volume} {79}},\ \bibinfo {pages} {241101} (\bibinfo {year}
  {2009})}\BibitemShut {NoStop}%
\bibitem [{\citenamefont {Liu}\ and\ \citenamefont
  {Allen}(1995)}]{liu_electronic_1995}%
  \BibitemOpen
  \bibfield  {author} {\bibinfo {author} {\bibfnamefont {Y.}~\bibnamefont
  {Liu}}\ and\ \bibinfo {author} {\bibfnamefont {R.~E.}\ \bibnamefont
  {Allen}},\ }\href {\doibase 10.1103/PhysRevB.52.1566} {\bibfield  {journal}
  {\bibinfo  {journal} {Physical Review B}\ }\textbf {\bibinfo {volume} {52}},\
  \bibinfo {pages} {1566} (\bibinfo {year} {1995})}\BibitemShut {NoStop}%
\bibitem [{\citenamefont {Zhang}\ \emph {et~al.}(2009)\citenamefont {Zhang},
  \citenamefont {Liu}, \citenamefont {Qi}, \citenamefont {Dai}, \citenamefont
  {Fang},\ and\ \citenamefont {Zhang}}]{zhang_Bi2Se3}%
  \BibitemOpen
  \bibfield  {author} {\bibinfo {author} {\bibfnamefont {H.}~\bibnamefont
  {Zhang}}, \bibinfo {author} {\bibfnamefont {C.-X.}\ \bibnamefont {Liu}},
  \bibinfo {author} {\bibfnamefont {X.-L.}\ \bibnamefont {Qi}}, \bibinfo
  {author} {\bibfnamefont {X.}~\bibnamefont {Dai}}, \bibinfo {author}
  {\bibfnamefont {Z.}~\bibnamefont {Fang}}, \ and\ \bibinfo {author}
  {\bibfnamefont {S.-C.}\ \bibnamefont {Zhang}},\ }\href {\doibase
  10.1038/nphys1270} {\bibfield  {journal} {\bibinfo  {journal} {Nature
  Physics}\ }\textbf {\bibinfo {volume} {5}},\ \bibinfo {pages} {438} (\bibinfo
  {year} {2009})}\BibitemShut {NoStop}%
\bibitem [{\citenamefont {Zhang}\ \emph {et~al.}(2010)\citenamefont {Zhang},
  \citenamefont {He}, \citenamefont {Chang}, \citenamefont {Song},
  \citenamefont {Wang}, \citenamefont {Chen}, \citenamefont {Jia},
  \citenamefont {Fang}, \citenamefont {Dai}, \citenamefont {Shan},
  \citenamefont {Shen}, \citenamefont {Niu}, \citenamefont {Qi}, \citenamefont
  {Zhang}, \citenamefont {Ma},\ and\ \citenamefont {Xue}}]{xueqk_Bi2Se3}%
  \BibitemOpen
  \bibfield  {author} {\bibinfo {author} {\bibfnamefont {Y.}~\bibnamefont
  {Zhang}}, \bibinfo {author} {\bibfnamefont {K.}~\bibnamefont {He}}, \bibinfo
  {author} {\bibfnamefont {C.-Z.}\ \bibnamefont {Chang}}, \bibinfo {author}
  {\bibfnamefont {C.-L.}\ \bibnamefont {Song}}, \bibinfo {author}
  {\bibfnamefont {L.-L.}\ \bibnamefont {Wang}}, \bibinfo {author}
  {\bibfnamefont {X.}~\bibnamefont {Chen}}, \bibinfo {author} {\bibfnamefont
  {J.-F.}\ \bibnamefont {Jia}}, \bibinfo {author} {\bibfnamefont
  {Z.}~\bibnamefont {Fang}}, \bibinfo {author} {\bibfnamefont {X.}~\bibnamefont
  {Dai}}, \bibinfo {author} {\bibfnamefont {W.-Y.}\ \bibnamefont {Shan}},
  \bibinfo {author} {\bibfnamefont {S.-Q.}\ \bibnamefont {Shen}}, \bibinfo
  {author} {\bibfnamefont {Q.}~\bibnamefont {Niu}}, \bibinfo {author}
  {\bibfnamefont {X.-L.}\ \bibnamefont {Qi}}, \bibinfo {author} {\bibfnamefont
  {S.-C.}\ \bibnamefont {Zhang}}, \bibinfo {author} {\bibfnamefont {X.-C.}\
  \bibnamefont {Ma}}, \ and\ \bibinfo {author} {\bibfnamefont {Q.-K.}\
  \bibnamefont {Xue}},\ }\href {\doibase 10.1038/nphys1689} {\bibfield
  {journal} {\bibinfo  {journal} {Nature Physics}\ }\textbf {\bibinfo {volume}
  {6}},\ \bibinfo {pages} {584} (\bibinfo {year} {2010})}\BibitemShut {NoStop}%
\bibitem [{\citenamefont {Altmann}\ and\ \citenamefont
  {Herzig}(1994)}]{altmann_point-group}%
  \BibitemOpen
  \bibfield  {author} {\bibinfo {author} {\bibfnamefont {S.~L.}\ \bibnamefont
  {Altmann}}\ and\ \bibinfo {author} {\bibfnamefont {P.}~\bibnamefont
  {Herzig}},\ }\href@noop {} {\emph {\bibinfo {title} {Point-group theory
  tables}}}\ (\bibinfo  {publisher} {Clarendon Press},\ \bibinfo {year}
  {1994})\BibitemShut {NoStop}%
\bibitem [{\citenamefont {Wang}\ \emph {et~al.}(2012)\citenamefont {Wang},
  \citenamefont {Sun}, \citenamefont {Chen}, \citenamefont {Franchini},
  \citenamefont {Xu}, \citenamefont {Weng}, \citenamefont {Dai},\ and\
  \citenamefont {Fang}}]{wang_Na3Bi}%
  \BibitemOpen
  \bibfield  {author} {\bibinfo {author} {\bibfnamefont {Z.}~\bibnamefont
  {Wang}}, \bibinfo {author} {\bibfnamefont {Y.}~\bibnamefont {Sun}}, \bibinfo
  {author} {\bibfnamefont {X.-Q.}\ \bibnamefont {Chen}}, \bibinfo {author}
  {\bibfnamefont {C.}~\bibnamefont {Franchini}}, \bibinfo {author}
  {\bibfnamefont {G.}~\bibnamefont {Xu}}, \bibinfo {author} {\bibfnamefont
  {H.}~\bibnamefont {Weng}}, \bibinfo {author} {\bibfnamefont {X.}~\bibnamefont
  {Dai}}, \ and\ \bibinfo {author} {\bibfnamefont {Z.}~\bibnamefont {Fang}},\
  }\href {\doibase 10.1103/PhysRevB.85.195320} {\bibfield  {journal} {\bibinfo
  {journal} {Physical Review B}\ }\textbf {\bibinfo {volume} {85}},\ \bibinfo
  {pages} {195320} (\bibinfo {year} {2012})}\BibitemShut {NoStop}%
\bibitem [{\citenamefont {Liu}\ \emph {et~al.}(2014)\citenamefont {Liu},
  \citenamefont {Zhou}, \citenamefont {Zhang}, \citenamefont {Wang},
  \citenamefont {Weng}, \citenamefont {Prabhakaran}, \citenamefont {Mo},
  \citenamefont {Shen}, \citenamefont {Fang}, \citenamefont {Dai},
  \citenamefont {Hussain},\ and\ \citenamefont {Chen}}]{Na3Bi_exp}%
  \BibitemOpen
  \bibfield  {author} {\bibinfo {author} {\bibfnamefont {Z.~K.}\ \bibnamefont
  {Liu}}, \bibinfo {author} {\bibfnamefont {B.}~\bibnamefont {Zhou}}, \bibinfo
  {author} {\bibfnamefont {Y.}~\bibnamefont {Zhang}}, \bibinfo {author}
  {\bibfnamefont {Z.~J.}\ \bibnamefont {Wang}}, \bibinfo {author}
  {\bibfnamefont {H.~M.}\ \bibnamefont {Weng}}, \bibinfo {author}
  {\bibfnamefont {D.}~\bibnamefont {Prabhakaran}}, \bibinfo {author}
  {\bibfnamefont {S.-K.}\ \bibnamefont {Mo}}, \bibinfo {author} {\bibfnamefont
  {Z.~X.}\ \bibnamefont {Shen}}, \bibinfo {author} {\bibfnamefont
  {Z.}~\bibnamefont {Fang}}, \bibinfo {author} {\bibfnamefont {X.}~\bibnamefont
  {Dai}}, \bibinfo {author} {\bibfnamefont {Z.}~\bibnamefont {Hussain}}, \ and\
  \bibinfo {author} {\bibfnamefont {Y.~L.}\ \bibnamefont {Chen}},\ }\href
  {\doibase 10.1126/science.1245085} {\bibfield  {journal} {\bibinfo  {journal}
  {Science}\ }\textbf {\bibinfo {volume} {343}},\ \bibinfo {pages} {864}
  (\bibinfo {year} {2014})}\BibitemShut {NoStop}%
\bibitem [{\citenamefont {Weng}\ \emph {et~al.}(2016)\citenamefont {Weng},
  \citenamefont {Fang}, \citenamefont {Fang},\ and\ \citenamefont
  {Dai}}]{weng_TaN}%
  \BibitemOpen
  \bibfield  {author} {\bibinfo {author} {\bibfnamefont {H.}~\bibnamefont
  {Weng}}, \bibinfo {author} {\bibfnamefont {C.}~\bibnamefont {Fang}}, \bibinfo
  {author} {\bibfnamefont {Z.}~\bibnamefont {Fang}}, \ and\ \bibinfo {author}
  {\bibfnamefont {X.}~\bibnamefont {Dai}},\ }\href {\doibase
  10.1103/PhysRevB.93.241202} {\bibfield  {journal} {\bibinfo  {journal}
  {Physical Review B}\ }\textbf {\bibinfo {volume} {93}},\ \bibinfo {pages}
  {241202} (\bibinfo {year} {2016})}\BibitemShut {NoStop}%
\bibitem [{\citenamefont {Weng}\ \emph {et~al.}(2014)\citenamefont {Weng},
  \citenamefont {Dai},\ and\ \citenamefont {Fang}}]{weng_ZrTe5}%
  \BibitemOpen
  \bibfield  {author} {\bibinfo {author} {\bibfnamefont {H.}~\bibnamefont
  {Weng}}, \bibinfo {author} {\bibfnamefont {X.}~\bibnamefont {Dai}}, \ and\
  \bibinfo {author} {\bibfnamefont {Z.}~\bibnamefont {Fang}},\ }\href {\doibase
  10.1103/PhysRevX.4.011002} {\bibfield  {journal} {\bibinfo  {journal}
  {Physical Review X}\ }\textbf {\bibinfo {volume} {4}},\ \bibinfo {pages}
  {011002} (\bibinfo {year} {2014})}\BibitemShut {NoStop}%
\bibitem [{\citenamefont {Fan}\ \emph {et~al.}(2017)\citenamefont {Fan},
  \citenamefont {Liang}, \citenamefont {Chen}, \citenamefont {Yao},\ and\
  \citenamefont {Zhou}}]{Zhouj_ZrTe5}%
  \BibitemOpen
  \bibfield  {author} {\bibinfo {author} {\bibfnamefont {Z.}~\bibnamefont
  {Fan}}, \bibinfo {author} {\bibfnamefont {Q.-F.}\ \bibnamefont {Liang}},
  \bibinfo {author} {\bibfnamefont {Y.~B.}\ \bibnamefont {Chen}}, \bibinfo
  {author} {\bibfnamefont {S.-H.}\ \bibnamefont {Yao}}, \ and\ \bibinfo
  {author} {\bibfnamefont {J.}~\bibnamefont {Zhou}},\ }\href {\doibase
  10.1038/srep45667} {\bibfield  {journal} {\bibinfo  {journal} {Scientific
  Reports}\ }\textbf {\bibinfo {volume} {7}} (\bibinfo {year} {2017}),\
  10.1038/srep45667}\BibitemShut {NoStop}%
\bibitem [{\citenamefont {Chen}\ \emph {et~al.}(2015)\citenamefont {Chen},
  \citenamefont {Chen}, \citenamefont {Song}, \citenamefont {Schneeloch},
  \citenamefont {Gu}, \citenamefont {Wang},\ and\ \citenamefont
  {Wang}}]{ZrTe5_massless_prl}%
  \BibitemOpen
  \bibfield  {author} {\bibinfo {author} {\bibfnamefont {R.~Y.}\ \bibnamefont
  {Chen}}, \bibinfo {author} {\bibfnamefont {Z.~G.}\ \bibnamefont {Chen}},
  \bibinfo {author} {\bibfnamefont {X.-Y.}\ \bibnamefont {Song}}, \bibinfo
  {author} {\bibfnamefont {J.~A.}\ \bibnamefont {Schneeloch}}, \bibinfo
  {author} {\bibfnamefont {G.~D.}\ \bibnamefont {Gu}}, \bibinfo {author}
  {\bibfnamefont {F.}~\bibnamefont {Wang}}, \ and\ \bibinfo {author}
  {\bibfnamefont {N.~L.}\ \bibnamefont {Wang}},\ }\href {\doibase
  10.1103/PhysRevLett.115.176404} {\bibfield  {journal} {\bibinfo  {journal}
  {Phys. Rev. Lett.}\ }\textbf {\bibinfo {volume} {115}},\ \bibinfo {pages}
  {176404} (\bibinfo {year} {2015})}\BibitemShut {NoStop}%
\bibitem [{\citenamefont {Yuan}\ \emph {et~al.}(2016)\citenamefont {Yuan},
  \citenamefont {Zhang}, \citenamefont {Liu}, \citenamefont {Narayan},
  \citenamefont {Song}, \citenamefont {Shen}, \citenamefont {Sui},
  \citenamefont {Xu}, \citenamefont {Yu}, \citenamefont {An} \emph
  {et~al.}}]{ZrTe5_2dDirac_NPG}%
  \BibitemOpen
  \bibfield  {author} {\bibinfo {author} {\bibfnamefont {X.}~\bibnamefont
  {Yuan}}, \bibinfo {author} {\bibfnamefont {C.}~\bibnamefont {Zhang}},
  \bibinfo {author} {\bibfnamefont {Y.}~\bibnamefont {Liu}}, \bibinfo {author}
  {\bibfnamefont {A.}~\bibnamefont {Narayan}}, \bibinfo {author} {\bibfnamefont
  {C.}~\bibnamefont {Song}}, \bibinfo {author} {\bibfnamefont {S.}~\bibnamefont
  {Shen}}, \bibinfo {author} {\bibfnamefont {X.}~\bibnamefont {Sui}}, \bibinfo
  {author} {\bibfnamefont {J.}~\bibnamefont {Xu}}, \bibinfo {author}
  {\bibfnamefont {H.}~\bibnamefont {Yu}}, \bibinfo {author} {\bibfnamefont
  {Z.}~\bibnamefont {An}},  \emph {et~al.},\ }\href@noop {} {\bibfield
  {journal} {\bibinfo  {journal} {NPG Asia Materials}\ }\textbf {\bibinfo
  {volume} {8}},\ \bibinfo {pages} {e325} (\bibinfo {year} {2016})}\BibitemShut
  {NoStop}%
\bibitem [{\citenamefont {Li}\ \emph {et~al.}(2016{\natexlab{a}})\citenamefont
  {Li}, \citenamefont {Kharzeev}, \citenamefont {Zhang}, \citenamefont {Huang},
  \citenamefont {Pletikosić}, \citenamefont {Fedorov}, \citenamefont {Zhong},
  \citenamefont {Schneeloch}, \citenamefont {Gu},\ and\ \citenamefont
  {Valla}}]{ZrTe5_CME}%
  \BibitemOpen
  \bibfield  {author} {\bibinfo {author} {\bibfnamefont {Q.}~\bibnamefont
  {Li}}, \bibinfo {author} {\bibfnamefont {D.~E.}\ \bibnamefont {Kharzeev}},
  \bibinfo {author} {\bibfnamefont {C.}~\bibnamefont {Zhang}}, \bibinfo
  {author} {\bibfnamefont {Y.}~\bibnamefont {Huang}}, \bibinfo {author}
  {\bibfnamefont {I.}~\bibnamefont {Pletikosić}}, \bibinfo {author}
  {\bibfnamefont {A.~V.}\ \bibnamefont {Fedorov}}, \bibinfo {author}
  {\bibfnamefont {R.~D.}\ \bibnamefont {Zhong}}, \bibinfo {author}
  {\bibfnamefont {J.~A.}\ \bibnamefont {Schneeloch}}, \bibinfo {author}
  {\bibfnamefont {G.~D.}\ \bibnamefont {Gu}}, \ and\ \bibinfo {author}
  {\bibfnamefont {T.}~\bibnamefont {Valla}},\ }\href {\doibase
  10.1038/nphys3648} {\bibfield  {journal} {\bibinfo  {journal} {Nature
  Physics}\ }\textbf {\bibinfo {volume} {advance online publication}} (\bibinfo
  {year} {2016}{\natexlab{a}}),\ 10.1038/nphys3648}\BibitemShut {NoStop}%
\bibitem [{\citenamefont {{Zhang}}\ \emph {et~al.}(2016)\citenamefont
  {{Zhang}}, \citenamefont {{Wang}}, \citenamefont {{Yu}}, \citenamefont
  {{Liu}}, \citenamefont {{Liang}}, \citenamefont {{Huang}}, \citenamefont
  {{Nie}}, \citenamefont {{Zhang}}, \citenamefont {{Shen}}, \citenamefont
  {{Liu}}, \citenamefont {{Weng}}, \citenamefont {{Zhao}}, \citenamefont
  {{Chen}}, \citenamefont {{Jia}}, \citenamefont {{Hu}}, \citenamefont
  {{Ding}}, \citenamefont {{He}}, \citenamefont {{Zhao}}, \citenamefont
  {{Zhang}}, \citenamefont {{Zhang}}, \citenamefont {{Yang}}, \citenamefont
  {{Wang}}, \citenamefont {{Peng}}, \citenamefont {{Dai}}, \citenamefont
  {{Fang}}, \citenamefont {{Xu}}, \citenamefont {{Chen}},\ and\ \citenamefont
  {{Zhou}}}]{ZrTe5_PT}%
  \BibitemOpen
  \bibfield  {author} {\bibinfo {author} {\bibfnamefont {Y.}~\bibnamefont
  {{Zhang}}}, \bibinfo {author} {\bibfnamefont {C.}~\bibnamefont {{Wang}}},
  \bibinfo {author} {\bibfnamefont {L.}~\bibnamefont {{Yu}}}, \bibinfo {author}
  {\bibfnamefont {G.}~\bibnamefont {{Liu}}}, \bibinfo {author} {\bibfnamefont
  {A.}~\bibnamefont {{Liang}}}, \bibinfo {author} {\bibfnamefont
  {J.}~\bibnamefont {{Huang}}}, \bibinfo {author} {\bibfnamefont
  {S.}~\bibnamefont {{Nie}}}, \bibinfo {author} {\bibfnamefont
  {Y.}~\bibnamefont {{Zhang}}}, \bibinfo {author} {\bibfnamefont
  {B.}~\bibnamefont {{Shen}}}, \bibinfo {author} {\bibfnamefont
  {J.}~\bibnamefont {{Liu}}}, \bibinfo {author} {\bibfnamefont
  {H.}~\bibnamefont {{Weng}}}, \bibinfo {author} {\bibfnamefont
  {L.}~\bibnamefont {{Zhao}}}, \bibinfo {author} {\bibfnamefont
  {G.}~\bibnamefont {{Chen}}}, \bibinfo {author} {\bibfnamefont
  {X.}~\bibnamefont {{Jia}}}, \bibinfo {author} {\bibfnamefont
  {C.}~\bibnamefont {{Hu}}}, \bibinfo {author} {\bibfnamefont {Y.}~\bibnamefont
  {{Ding}}}, \bibinfo {author} {\bibfnamefont {S.}~\bibnamefont {{He}}},
  \bibinfo {author} {\bibfnamefont {L.}~\bibnamefont {{Zhao}}}, \bibinfo
  {author} {\bibfnamefont {F.}~\bibnamefont {{Zhang}}}, \bibinfo {author}
  {\bibfnamefont {S.}~\bibnamefont {{Zhang}}}, \bibinfo {author} {\bibfnamefont
  {F.}~\bibnamefont {{Yang}}}, \bibinfo {author} {\bibfnamefont
  {Z.}~\bibnamefont {{Wang}}}, \bibinfo {author} {\bibfnamefont
  {Q.}~\bibnamefont {{Peng}}}, \bibinfo {author} {\bibfnamefont
  {X.}~\bibnamefont {{Dai}}}, \bibinfo {author} {\bibfnamefont
  {Z.}~\bibnamefont {{Fang}}}, \bibinfo {author} {\bibfnamefont
  {Z.}~\bibnamefont {{Xu}}}, \bibinfo {author} {\bibfnamefont {C.}~\bibnamefont
  {{Chen}}}, \ and\ \bibinfo {author} {\bibfnamefont {X.~J.}\ \bibnamefont
  {{Zhou}}},\ }\href@noop {} {\bibfield  {journal} {\bibinfo  {journal} {ArXiv
  e-prints}\ } (\bibinfo {year} {2016})},\ \Eprint
  {http://arxiv.org/abs/1602.03576} {arXiv:1602.03576 [cond-mat.mtrl-sci]}
  \BibitemShut {NoStop}%
\bibitem [{\citenamefont {Manzoni}\ \emph {et~al.}(2016)\citenamefont
  {Manzoni}, \citenamefont {Gragnaniello}, \citenamefont {Aut\`es},
  \citenamefont {Kuhn}, \citenamefont {Sterzi}, \citenamefont {Cilento},
  \citenamefont {Zacchigna}, \citenamefont {Enenkel}, \citenamefont {Vobornik},
  \citenamefont {Barba}, \citenamefont {Bisti}, \citenamefont {Bugnon},
  \citenamefont {Magrez}, \citenamefont {Strocov}, \citenamefont {Berger},
  \citenamefont {Yazyev}, \citenamefont {Fonin}, \citenamefont {Parmigiani},\
  and\ \citenamefont {Crepaldi}}]{ZrTe5_strong_prl}%
  \BibitemOpen
  \bibfield  {author} {\bibinfo {author} {\bibfnamefont {G.}~\bibnamefont
  {Manzoni}}, \bibinfo {author} {\bibfnamefont {L.}~\bibnamefont
  {Gragnaniello}}, \bibinfo {author} {\bibfnamefont {G.}~\bibnamefont
  {Aut\`es}}, \bibinfo {author} {\bibfnamefont {T.}~\bibnamefont {Kuhn}},
  \bibinfo {author} {\bibfnamefont {A.}~\bibnamefont {Sterzi}}, \bibinfo
  {author} {\bibfnamefont {F.}~\bibnamefont {Cilento}}, \bibinfo {author}
  {\bibfnamefont {M.}~\bibnamefont {Zacchigna}}, \bibinfo {author}
  {\bibfnamefont {V.}~\bibnamefont {Enenkel}}, \bibinfo {author} {\bibfnamefont
  {I.}~\bibnamefont {Vobornik}}, \bibinfo {author} {\bibfnamefont
  {L.}~\bibnamefont {Barba}}, \bibinfo {author} {\bibfnamefont
  {F.}~\bibnamefont {Bisti}}, \bibinfo {author} {\bibfnamefont
  {P.}~\bibnamefont {Bugnon}}, \bibinfo {author} {\bibfnamefont
  {A.}~\bibnamefont {Magrez}}, \bibinfo {author} {\bibfnamefont {V.~N.}\
  \bibnamefont {Strocov}}, \bibinfo {author} {\bibfnamefont {H.}~\bibnamefont
  {Berger}}, \bibinfo {author} {\bibfnamefont {O.~V.}\ \bibnamefont {Yazyev}},
  \bibinfo {author} {\bibfnamefont {M.}~\bibnamefont {Fonin}}, \bibinfo
  {author} {\bibfnamefont {F.}~\bibnamefont {Parmigiani}}, \ and\ \bibinfo
  {author} {\bibfnamefont {A.}~\bibnamefont {Crepaldi}},\ }\href {\doibase
  10.1103/PhysRevLett.117.237601} {\bibfield  {journal} {\bibinfo  {journal}
  {Phys. Rev. Lett.}\ }\textbf {\bibinfo {volume} {117}},\ \bibinfo {pages}
  {237601} (\bibinfo {year} {2016})}\BibitemShut {NoStop}%
\bibitem [{\citenamefont {Li}\ \emph {et~al.}(2016{\natexlab{b}})\citenamefont
  {Li}, \citenamefont {Huang}, \citenamefont {Lv}, \citenamefont {Zhang},
  \citenamefont {Yang}, \citenamefont {Zhang}, \citenamefont {Chen},
  \citenamefont {Yao}, \citenamefont {Zhou}, \citenamefont {Lu}, \citenamefont
  {Sheng}, \citenamefont {Li}, \citenamefont {Jia}, \citenamefont {Xue},
  \citenamefont {Chen},\ and\ \citenamefont {Xing}}]{ZrTe5_weak_prl}%
  \BibitemOpen
  \bibfield  {author} {\bibinfo {author} {\bibfnamefont {X.-B.}\ \bibnamefont
  {Li}}, \bibinfo {author} {\bibfnamefont {W.-K.}\ \bibnamefont {Huang}},
  \bibinfo {author} {\bibfnamefont {Y.-Y.}\ \bibnamefont {Lv}}, \bibinfo
  {author} {\bibfnamefont {K.-W.}\ \bibnamefont {Zhang}}, \bibinfo {author}
  {\bibfnamefont {C.-L.}\ \bibnamefont {Yang}}, \bibinfo {author}
  {\bibfnamefont {B.-B.}\ \bibnamefont {Zhang}}, \bibinfo {author}
  {\bibfnamefont {Y.~B.}\ \bibnamefont {Chen}}, \bibinfo {author}
  {\bibfnamefont {S.-H.}\ \bibnamefont {Yao}}, \bibinfo {author} {\bibfnamefont
  {J.}~\bibnamefont {Zhou}}, \bibinfo {author} {\bibfnamefont {M.-H.}\
  \bibnamefont {Lu}}, \bibinfo {author} {\bibfnamefont {L.}~\bibnamefont
  {Sheng}}, \bibinfo {author} {\bibfnamefont {S.-C.}\ \bibnamefont {Li}},
  \bibinfo {author} {\bibfnamefont {J.-F.}\ \bibnamefont {Jia}}, \bibinfo
  {author} {\bibfnamefont {Q.-K.}\ \bibnamefont {Xue}}, \bibinfo {author}
  {\bibfnamefont {Y.-F.}\ \bibnamefont {Chen}}, \ and\ \bibinfo {author}
  {\bibfnamefont {D.-Y.}\ \bibnamefont {Xing}},\ }\href {\doibase
  10.1103/PhysRevLett.116.176803} {\bibfield  {journal} {\bibinfo  {journal}
  {Phys. Rev. Lett.}\ }\textbf {\bibinfo {volume} {116}},\ \bibinfo {pages}
  {176803} (\bibinfo {year} {2016}{\natexlab{b}})}\BibitemShut {NoStop}%
\bibitem [{\citenamefont {Wu}\ \emph {et~al.}(2016)\citenamefont {Wu},
  \citenamefont {Ma}, \citenamefont {Nie}, \citenamefont {Zhao}, \citenamefont
  {Huang}, \citenamefont {Yin}, \citenamefont {Fu}, \citenamefont {Richard},
  \citenamefont {Chen}, \citenamefont {Fang}, \citenamefont {Dai},
  \citenamefont {Weng}, \citenamefont {Qian}, \citenamefont {Ding},\ and\
  \citenamefont {Pan}}]{ZrTe5_weak_prx}%
  \BibitemOpen
  \bibfield  {author} {\bibinfo {author} {\bibfnamefont {R.}~\bibnamefont
  {Wu}}, \bibinfo {author} {\bibfnamefont {J.-Z.}\ \bibnamefont {Ma}}, \bibinfo
  {author} {\bibfnamefont {S.-M.}\ \bibnamefont {Nie}}, \bibinfo {author}
  {\bibfnamefont {L.-X.}\ \bibnamefont {Zhao}}, \bibinfo {author}
  {\bibfnamefont {X.}~\bibnamefont {Huang}}, \bibinfo {author} {\bibfnamefont
  {J.-X.}\ \bibnamefont {Yin}}, \bibinfo {author} {\bibfnamefont {B.-B.}\
  \bibnamefont {Fu}}, \bibinfo {author} {\bibfnamefont {P.}~\bibnamefont
  {Richard}}, \bibinfo {author} {\bibfnamefont {G.-F.}\ \bibnamefont {Chen}},
  \bibinfo {author} {\bibfnamefont {Z.}~\bibnamefont {Fang}}, \bibinfo {author}
  {\bibfnamefont {X.}~\bibnamefont {Dai}}, \bibinfo {author} {\bibfnamefont
  {H.-M.}\ \bibnamefont {Weng}}, \bibinfo {author} {\bibfnamefont
  {T.}~\bibnamefont {Qian}}, \bibinfo {author} {\bibfnamefont {H.}~\bibnamefont
  {Ding}}, \ and\ \bibinfo {author} {\bibfnamefont {S.~H.}\ \bibnamefont
  {Pan}},\ }\href {\doibase 10.1103/PhysRevX.6.021017} {\bibfield  {journal}
  {\bibinfo  {journal} {Phys. Rev. X}\ }\textbf {\bibinfo {volume} {6}},\
  \bibinfo {pages} {021017} (\bibinfo {year} {2016})}\BibitemShut {NoStop}%
\bibitem [{\citenamefont {Liang}\ \emph {et~al.}(2016)\citenamefont {Liang},
  \citenamefont {Gibson}, \citenamefont {Liu}, \citenamefont {Wang},
  \citenamefont {Cava},\ and\ \citenamefont {Ong}}]{ZrTe5_Hall}%
  \BibitemOpen
  \bibfield  {author} {\bibinfo {author} {\bibfnamefont {T.}~\bibnamefont
  {Liang}}, \bibinfo {author} {\bibfnamefont {Q.}~\bibnamefont {Gibson}},
  \bibinfo {author} {\bibfnamefont {M.}~\bibnamefont {Liu}}, \bibinfo {author}
  {\bibfnamefont {W.}~\bibnamefont {Wang}}, \bibinfo {author} {\bibfnamefont
  {R.~J.}\ \bibnamefont {Cava}}, \ and\ \bibinfo {author} {\bibfnamefont
  {N.~P.}\ \bibnamefont {Ong}},\ }\href {http://arxiv.org/abs/1612.06972}
  {\bibfield  {journal} {\bibinfo  {journal} {arXiv:1612.06972 [cond-mat]}\ }
  (\bibinfo {year} {2016})},\ \bibinfo {note} {arXiv: 1612.06972}\BibitemShut
  {NoStop}%
\bibitem [{\citenamefont {Blöchl}(1994)}]{blochl_PAW}%
  \BibitemOpen
  \bibfield  {author} {\bibinfo {author} {\bibfnamefont {P.~E.}\ \bibnamefont
  {Blöchl}},\ }\href {\doibase 10.1103/PhysRevB.50.17953} {\bibfield
  {journal} {\bibinfo  {journal} {Physical Review B}\ }\textbf {\bibinfo
  {volume} {50}},\ \bibinfo {pages} {17953} (\bibinfo {year}
  {1994})}\BibitemShut {NoStop}%
\bibitem [{\citenamefont {Kresse}\ and\ \citenamefont
  {Joubert}(1999)}]{kresse_PAW}%
  \BibitemOpen
  \bibfield  {author} {\bibinfo {author} {\bibfnamefont {G.}~\bibnamefont
  {Kresse}}\ and\ \bibinfo {author} {\bibfnamefont {D.}~\bibnamefont
  {Joubert}},\ }\href {\doibase 10.1103/PhysRevB.59.1758} {\bibfield  {journal}
  {\bibinfo  {journal} {Physical Review B}\ }\textbf {\bibinfo {volume} {59}},\
  \bibinfo {pages} {1758} (\bibinfo {year} {1999})}\BibitemShut {NoStop}%
\end{thebibliography}%



\end{document}